\documentclass[aps,nofootinbib,notitlepage,superscriptaddress,twocolumn,10pt,prd]{revtex4-2}

\usepackage{bm}
\usepackage{graphicx}
\usepackage{amsmath,amssymb}
\usepackage{hyperref}
\usepackage{braket}
\usepackage{subfigure}
\usepackage{float}
\usepackage[dvipsnames]{xcolor}
\usepackage{soul}
\usepackage{rotating}
\usepackage{multirow}
\usepackage{mathtools}
\usepackage{lipsum}
\usepackage{makecell}

\usepackage[margin=0.75in]{geometry}

\hypersetup{
	colorlinks=true,
	linkcolor=red,
	citecolor=blue,
}
\usepackage{physics}

\usepackage{natbib}
\usepackage{verbatim}
\usepackage{dcolumn}
\usepackage[amssymb]{SIunits}
\usepackage{tabularx}
\usepackage{booktabs}
\usepackage[normalem]{ulem}

\newcommand{\be}{\begin{equation}}
\newcommand{\ee}{\end{equation}}
\newcommand{\ba}{\begin{eqnarray}}
\newcommand{\ea}{\end{eqnarray}}

\definecolor{ForestGreen}{RGB}{36,179,0}

\begin{document}

\title{Beyond dark energy Fisher forecasts: how DESI will constrain LCDM and quintessence models}

\author{Samuel Goldstein}
\affiliation{Center for Particle Cosmology, Department of Physics and Astronomy, University of Pennsylvania, Philadelphia, PA 19104, USA}
\affiliation{Max-Planck-Institute für Astrophysik,  Karl–Schwarzschild–Straße 1, 85748 Garching, Germany}
\author{Minsu Park}
\affiliation{Center for Particle Cosmology, Department of Physics and Astronomy, University of Pennsylvania, Philadelphia, PA 19104, USA}
\author{Marco Raveri}
\affiliation{Center for Particle Cosmology, Department of Physics and Astronomy, University of Pennsylvania, Philadelphia, PA 19104, USA}
\affiliation{Department of Physics, 
University of Genova, Via Dodecaneso 33, 16146, Italy}
\author{Bhuvnesh Jain}
\affiliation{Center for Particle Cosmology, Department of Physics and Astronomy, University of Pennsylvania, Philadelphia, PA 19104, USA}
\author{Lado Samushia}
\affiliation{Department of Physics, Kansas State University, 116 Cardwell Hall, Manhattan, KS 66506, USA}
\affiliation{Abastumani Astrophysical Observatory, Tbilisi, GE-0179, Georgia}

\begin{abstract}

We baseline with current cosmological observations to forecast the power of the Dark Energy Spectroscopic Instrument (DESI) in two ways: 1. the gain in constraining power of parameter combinations in the standard $\Lambda$CDM model, and 2. the reconstruction of quintessence models of dark energy. For the former task we use a recently developed formalism to extract the leading parameter combinations constrained by different combinations of cosmological survey data. For the latter, we perform a non-parametric reconstruction of quintessence using the Effective Field Theory of Dark Energy. Using mock DESI observations of the Hubble parameter, angular diameter distance, and growth rate, we find that DESI will provide significant improvements over current datasets on $\Lambda$CDM and quintessence constraints. Including DESI mocks in our $\Lambda$CDM analysis improves constraints on $\Omega_m$, $H_0$, and $\sigma_8$ by a factor of two, where the improvement results almost entirely from the angular diameter distance and growth of structure measurements. Our quintessence reconstruction suggests that DESI will considerably improve constraints on a range of quintessence properties, such as the reconstructed potential, scalar field excursion, and the dark energy equation of state. The angular diameter distance measurements are particularly constraining in the presence of a non-$\Lambda$CDM signal in which the potential cannot be accounted for by shifts in $H_0$ and $\Omega_m$. 
\end{abstract}

\maketitle

\section{Introduction} \label{Sec:Intro}

Observations of the late time accelerated expansion of the Universe \cite{Riess:1998cb, Perlmutter:1998np} have motivated the introduction of a novel dark energy component to our Universe. A myriad of theories have been proposed to explain the nature of dark energy, with a cosmological constant providing the simplest and thus far observationally sound explanation. Nevertheless, the precise value of the cosmological constant appears to be heavily fine-tuned and is significantly smaller than expectations from particle physics \cite{weinberg1989, carroll}. These challenges motivate the exploration of alternative dark energy models, the simplest of which introduce new scalar degrees of freedom to our Universe.

Quintessence is a dark energy model in which accelerated expansion is sourced by a slowly varying canonical scalar field. Certain quintessence models can avoid the fine-tuning issues of the cosmological constant via the existence of tracking solutions \cite{Zlatev1999, Sahni}, although many of the models that possess such tracking solutions are inconsistent with observations \cite{bludman2001}. Recent studies have used a range of cosmological datasets to place constraints on quintessence \cite{Park:2021qnt, Yang_2019, Sangwan218}; however, on going and upcoming wide-field galaxy surveys such as the Dark Energy Spectroscopic Instrument (DESI) \cite{Aghamousa:2017}, the Legacy Survey of Space and Time (LSST) from the Vera Rubin Observatory \cite{LSST}, Euclid \cite{Euclid}, and the Nancy Grace Roman Space Telescope \cite{Roman} have the potential to improve these constraints considerably. One of the main goals of this work is to forecast this improvement for DESI.

DESI is a Stage IV spectroscopic galaxy and quasar survey designed to study the nature of dark energy. DESI will measure the redshifts of approximately 35 million galaxies and quasars and constrain the expansion rate via the baryon acoustic oscillations (BAO) and the growth of structure from the redshift-space distortions (RSD) with an unprecedented level of precision and redshift coverage \cite{Aghamousa:2017}. These observables are especially sensitive to the background evolution and structure growth of the Universe and hence DESI is an opportune experiment to test dark energy theories such as quintessence.

 In this work we investigate how DESI observations of the Hubble parameter, angular diameter distance, and linear growth rate will improve constraints on $\Lambda$CDM and quintessence. For the former, we utilize recently developed linear methods \cite{Dacunha2022} to extract the leading parameter combinations constrained by different combinations of cosmological survey data.  For the latter, we follow the approach of \citet{Park:2021qnt} and perform a non-parametric effective field theory based reconstruction of quintessence models using a combination of cosmic microwave background (CMB), large scale structure (LSS), supernova (SN), and mock DESI datasets. We find that including mock DESI measurements, particularly of the angular diameter distance and growth rate, considerably improves constraints on $\Lambda$CDM and quintessence cosmologies relative to current CMB, LSS, and SN datasets.
 
 This paper is organized as follows. In Sec.~\ref{Sec:Theory} we review quintessence and its mapping in the Effective Field Theory of Dark Energy. In Sec.~\ref{Sec:methodology} we discuss the EFT reconstruction, model fitting, and analysis methods. In Sec.~\ref{Sec:datasets_mocks} we discuss our dataset choice and mock DESI observables. We present our results in Sec.~\ref{Sec:Results} and conclusions in Sec.~\ref{Sec:Conclusions}.

\clearpage
\pagebreak

\section{Quintessence reconstruction} \label{Sec:Theory}
Quintessence is a dynamical dark energy model consisting of a canonical scalar field with potential energy $V(\phi)$ that is minimally coupled to gravity. The quintessence contribution to the action is given by
\begin{align} \label{Eq:quintessenceAction}
S_\phi \equiv \int d^4x \, \sqrt{-g} \left( \frac12 \nabla_\mu \phi \nabla^\mu \phi - V(\phi) \right) \,.
\end{align}
Combined with the Einstein-Hilbert action and the action for matter, we can derive the two Friedmann equations for quintessence, which fully describe the background evolution of a  Friedmann-Lemaître-Robertson-Walker (FLRW) universe:
\begin{align} \label{Eq:FRWequations}
& 3 M_P^2 \mathcal{H}^2 =  \frac{\dot{\phi}^2}{2} + V(\phi) a^2 + \rho_{\rm m} a^2 \,, \nonumber \\
& 6 M_P^2 \dot{\mathcal{H}} = -2 \dot{\phi}^2 +2 V(\phi) a^2 -(\rho_{\rm m} + 3 P_{\rm m} )a^2 \,, 
\end{align}
where $\mathcal{H} = d\ln a/d\tau$; the overdot is derivative with respect to conformal time; terms with a subscript ${\rm m}$ are summed over matter species; and Planck mass is denoted by $M_P^2 = 1 / (8\pi G_N)$. The equation of motion for $\phi$ is given by the Klein-Gordon equation:
\begin{align}
\ddot{\phi} +2 \mathcal{H} \dot{\phi} + a^2 \frac{\partial V}{\partial \phi} = 0 
\end{align}
which can be derived from Eqs.~\eqref{Eq:FRWequations}. 

Given the properties of matter species and the shape of $V(\phi)$, the history of the background evolution of the Universe is fully fixed by the boundary conditions of $\phi$ and $\dot{\phi}$. Notice that $V(\phi)$ with initial conditions $(\phi_0, \ \dot{\phi}_0)$ produces the same evolution history as $V(\phi-\delta\phi)$ with initial conditions $(\phi_0+\delta\phi, \dot{\phi}_0)$. That is, shifting the potential is degenerate with shifting the initial condition $\phi_0$. This symmetry makes it impossible to fully reconstruct $V(\phi)$ from a given evolution history. 

The Effective Field Theory of Dark Energy (EFTofDE) enables us to reconstruct the effects of quintessence on the background while remaining independent of the specific shape of $V(\phi)$~\cite{Bloomfield:2012ff, Gubitosi:2012hu, Raveri:2014cka}. In full generality, the EFTofDE encapsulates all possible deviations from the $\Lambda$CDM background assuming isotropy and homogeneity of the Universe. Quintessence is fully described in this framework by two free functions of time $\Lambda$ and $c$, whose Lagrangian is given by:
\begin{align} \label{Eq:EFTBackgroundAction}
S_{\Lambda,c} \equiv \int d^4x \sqrt{-g}  \left[ \Lambda(\tau) - c(\tau)\,a^2\delta g^{00} \right] \,,
\end{align}
The Friedmann equations of this action are then given by:
\begin{align} \label{Eq:EFTFRWequations}
3 M_P^2 \mathcal{H}^2 &= 2 c a^2 - \Lambda a^2 + \rho_{\rm m} a^2  \,, \nonumber \\
6 M_P^2 \dot{\mathcal{H}} &= -2(c+\Lambda) a^2 - (\rho_{\rm m} + 3 P_{\rm m}) a^2 \,.
\end{align}
where $c(\tau)$ can be eliminated from the equations to obtain a differential equation for $\mathcal{H}$ as a function of $\Lambda(z)$ only. This is consistent with the Lagrangian in Eq.~\eqref{Eq:quintessenceAction} which has only one non-constrained free function. Note though that we can always compute $c(z)$ from these equations, should we need it. Since there is a fiducial constant value for $\Lambda$ given the standard $\Lambda$CDM cosmology, it is convenient for later comparisons to consider the free function $\Lambda$  in terms of the relative difference to that fiducial constant at any given moment. That is, we define
\begin{align} \label{Eq:EFT_function}
\frac{\Delta\Lambda}{\Lambda} \equiv \frac{\Lambda}{3(1-\Omega_{\rm m})M_P^2\mathcal{H}_0^2} -1  \,
\end{align}
for our numerical algorithms and illustrative plots. 

One can see in both the action and the Friedmann equations that in a FLRW background Eqs.~\eqref{Eq:quintessenceAction} and~\eqref{Eq:EFTBackgroundAction} are equivalent given the mapping: 
\begin{align} \label{Eq:quintessenceMapping}
\Lambda =&\, \frac{1}{2a^2} \dot{\phi}^2 -V(\phi) \ \,, \nonumber \\
c =&\, \frac{1}{2a^2} \dot{\phi}^2 \,.
   \end{align}
This means that we can describe the physics of quintessence purely in terms of the EFTofDE. First, the equation of state for quintessence dark energy is given by: 
\begin{align} \label{Eq:wde}
w_{\rm DE} \equiv \frac{P_{\rm DE}}{\rho_{\rm DE}} \equiv \frac{\frac{1}{2a^2} \dot{\phi}^2 -V}{\frac{1}{2a^2} \dot{\phi}^2 + V }    = \frac{\Lambda}{2c-\Lambda} \,. 
\end{align}
Since $\frac{1}{2a^2} \dot{\phi}^2>0$, quintessence models must have $w_{\rm DE} \geq -1$. This is equivalent to saying that $c>0$~\cite{Vikman:2004dc, Hu:2004kh, Caldwell:2005ai, Creminelli:2008wc}.

From $w_{\rm DE}$ and its time derivative, quintessence can be said to be ``thawing'' and ``freezing''~\cite{Caldwell:2005tm,Scherrer:2005je, Linder:2006sv, Cortes:2009kc}. Here the useful time derivative of $w_{\rm DE}$ is $d w_{\rm DE}/d\ln a$. ``Thawing'' is a scenario where initially $\phi$ is kept stationary on a potential slope by the Hubble friction of the early universe. Then in the late universe the friction decreases and $\phi$ rolls towards the minimum, lifting $w_{\rm DE}$ above $-1$. ``Freezing'' is where initially $\phi$ is rolling towards the minimum, and the potential and Hubble friction slow it down near the minimum. $w_{\rm DE}$ moves towards $-1$ as time goes in this scenario.

Recently it has been conjectured that dark energy must be a non-negligibly dynamic scalar field given that a cosmological constant is yet to be realized with string theory~\cite{Ooguri:2018wrx, Agrawal:2018own, Garg:2018reu}. Quantitatively, these Swampland Conjectures place restrictions on properties of the potential and evolution of $\phi$ such that dark energy is sufficiently distinct from a cosmological constant. In Planck units, these bounds for a single field quintessence are: 
\begin{align} \label{Eq:SwamplandConj}
\abs{\nabla_\phi V}/V \gtrsim \mathcal{O}(1) \quad &\textbf{or} \quad -\nabla_\phi^2 V/V \gtrsim \mathcal{O}(1)  \nonumber \\
\abs{\Delta\phi} &\lesssim \mathcal{O}(1).
\end{align}
Roughly, for a positive $V(a)$, the first conjecture says that the potential must be sufficiently steep or the second derivative of the potential must be sufficiently negative at any given time. This rules out a positive cosmological constant. The second conjecture says that the field excursion ($\abs{\Delta\phi}$) must be limited to a radius of around one Planck mass. See e.g.~\cite{Akrami:2018ylq,Kinney:2018nny, Raveri:2018ddi} for further discussion. The first step in computing the quantities relevant to the Swampland Conjectures is determining the value of the potential energy as a function of time and $\dot{\phi}$:
\begin{align} \label{Eq:Va}
V(a) = c - \Lambda, \quad \dot{\phi}(a) = a\sqrt{2c} \, . 
\end{align}
We assume that in the period of interest, $\phi$ evolved monotonically with time. With $\Lambda$ as a function of $N=\ln a$, we can compute $\mathcal{H}$ as a function of $N$, from which we can compute $c$ and $\dot\phi$. Note the identity:
\begin{align} \label{eq:phidir}
\frac{d\phi}{d N} = \left(\frac{dN}{d\tau}\right)^{-1} \frac{d\phi}{d\tau}= \frac{\sqrt{2 c a^2}}{\mathcal{H}} \,.
\end{align}
We can then compute the quantities appearing in Eq.~\ref{Eq:SwamplandConj}
\begin{align}
\frac{\nabla_\phi V}{V} &= \frac1V \frac{dV/dN}{d\phi/dN} = \frac{d \ln V}{dN} \sqrt{\frac{\mathcal{H}^2}{2ca^2}} \,, \nonumber \\
\frac{\nabla_\phi^2 V}{V} &= \frac1V \left( \frac{\partial_N^2 V}{(\partial_N \phi)^2} - \frac{(\partial_N V)( \partial_N^2\phi ) }{(\partial_N\phi)^3}  \right) \,, \nonumber \\
\Delta\phi &= \int d\phi = \int_{0}^N \frac{\sqrt{2ca^2}}{\mathcal{H}}  \, dN \,,
\end{align}
where we chose to define $\Delta\phi$ with respect to the present day ($N=0$) since we have full freedom to shift $\phi$ (and therefore the freedom to chose an ``origin'' for $\phi$).

\section{Methodology} \label{Sec:methodology}
\subsection{EFT Reconstruction}\label{Sec:eft_recon}

In order to constrain the time-evolution of $\Delta \Lambda/\Lambda$ with minimal assumptions about its functional form we follow the methodology outlined in~\cite{Crittenden:2011aa, Raveri:2019mxg} in which one reconstructs a function of time by imposing a correlation prior on its temporal variations.  The correlation prior enforces the smoothness of the EFT function and acts as a low pass filter in order to avoid over fitting noise or systematics.

We reconstruct $\Delta \Lambda/\Lambda$ in the range $a\in [0.1,1.0]$ with a correlation length of $\Delta a=0.3$ following~\cite{Raveri:2019mxg}. For a fixed reconstruction interval and prior correlation length, the EFT function is represented by a piece-wise quintic spline passing through a sufficient number of nodes per correlation length. Our reconstruction range consists of three correlation lengths and we pick 15 spline nodes linearly spaced between $a=0.1$ and $a=1.0$, which provides five spline nodes per correlation length. Adding more spline nodes would not change our results as the correlation prior restricts sub-correlation length variations in the reconstructed function.

Since the correlation prior derived in \cite{Raveri:2019mxg} is only valid for late times, we restrict our reconstruction such that it recovers $\Lambda$CDM behavior at early times by imposing $\Delta \Lambda(a=0.1)/\Lambda=0.$  This restriction affects the reconstruction throughout the first correlation length ($a=0.4$), hence we only report results for later times, i.e. redshift $z<1.5.$

\subsection{Model Fitting}

We analyze our datasets with respect to both $\Lambda$CDM and quintessence cosmologies. In the $\Lambda$CDM analysis we vary the six cosmological parameters of the $\Lambda$CDM model ($\Omega_bh^2$, $\Omega_ch^2$, $A_s$, $n_s$, $\tau$, $\theta_s$) assuming priors from~\cite{Aghanim:2018eyx}. We also include all of the recommended nuisance parameters and priors to account for systematic effects in the datasets we consider. We fix the sum of neutrino masses to the minimal value of $0.06$ eV. In the quintessence analysis we include all parameters from the $\Lambda$CDM analysis, as well as 14 parameters consisting of values of $\Delta \Lambda/\Lambda$ at the spline nodes described in Sec.~\ref{Sec:eft_recon} with prior given by the correlation prior.

We use the EFTCAMB and EFTCosmoMC codes to sample the parameter posterior distributions~\cite{Hu:2013twa, Raveri:2014cka}. EFTCAMB and EFTCosmoMC are publicly available modifications of the Einstein-Boltzmann code CAMB and Markov Chain Monte Carlo (MCMC) code CosmoMC that incorporate the EFTofDE. We assess convergence using the Gelman-Rubin statistic~\cite{gelman1992} with a tolerance of $R-1<0.02.$

\subsection{Covariant Principal Component Analysis} \label{Sec:linear_cpca}
We use covariant principal component analysis (CPCA) to quantify the information gain and improvement on parameter constraints with the addition of mock DESI data. In this section, we provide an overview of CPCA. For a more detailed treatment see~\cite{Dacunha2022}.

We first establish some notation. Consider a model $\mathcal{M}$ with parameters $\theta$, and a measured dataset $D$. The key distribution of interest is the posterior, $\mathcal{P}(\theta)\equiv P(\theta \mid D, \mathcal{M})$, which is defined as the probability distribution of parameters given the data and the model. By Bayes theorem the posterior is given by
\begin{align}
\mathcal{P}(\theta\mid D, \mathcal{M})=\frac{\mathcal{L}(\theta)\Pi(\theta)}{\mathcal{E}} \,,
\end{align}
where $\Pi(\theta)\equiv P(\theta\mid \mathcal{M})$ is the prior distribution, $\mathcal{L}(\theta)\equiv P(\mathcal{D}\mid \theta, \mathcal{M})$ is the likelihood, and $\mathcal{E}\equiv P(D\mid \mathcal{M})=\int\mathcal{L}(\theta)\Pi(\theta)d\theta$ is the evidence. We let $\mathcal{C}_{\Pi}$ denote the prior covariance, $\mathcal{C}_{D}$ denote the data covariance, and $\mathcal{C}_{p_D}$ denote the posterior covariance associated with $D.$

Since we are interested in studying the improvement in the parameter constraints for a particular model resulting from adding new datasets, in what follows we consider a fixed model $\mathcal{M}$, prior $\Pi$, and datasets $A$, $B$, and their combination $A+B$. We wish to quantify the improvement in our constraining power from the addition of dataset $B$. This can be phrased as a generalized eigenvalue problem of the posterior covariances:
\begin{align} \label{eq:geneig}
\mathcal{C}_{p_{A+B}}^{-1}\Psi=\mathcal{C}_{p_{A}}^{-1}\Psi\Lambda \,,
\end{align}
where $\Psi$ are the CPCA modes and $\Lambda$ are the CPCA eigenvalues. The CPCA modes are the most improved linear combination of parameters after including dataset $B$. The CPCA eigenvalues can be written as $\lambda_i=\sigma^2_{p_{A,i}}/\sigma^2_{p_{A+B,i}}-1$ and hence quantify the improvement in the posterior covariance of the $i_{\rm th}$ CPCA mode after including dataset $B.$ There exists a unique solution to Eq.~\ref{eq:geneig} up to permutations given the constraints:
\begin{align}
\Psi^T\mathcal{C}_{p_{A}}^{-1}\Psi&=I \,, \\
\Psi^T\mathcal{C}_{p_{A+B}}^{-1}\Psi&=\Lambda \,.
\end{align}
The first constraint indicates that the posterior covariance conditioned on dataset $A$ establishes the units of our analysis. The second shows that $\Lambda$ quantifies the improvement in the posterior covariance after including dataset $B.$ 

Note that CPCA modes are consistent across invertible affine transformations of parameters unlike traditional principle component analyses (PCA) of covariance matrices. To illustrate, first consider the PCA as an eigenvalue problem: 
\begin{align} 
\mathcal{C}^{-1} \Psi = \Psi \Lambda\, .
\end{align}
Where $\mathcal{C}$ is a covariance matrix and $\Psi$ are the principle modes. Under affine transformations of parameters $\theta \mapsto A\theta$, covariance matrices transform as $\mathcal{C} \mapsto \mathcal{C}^\prime \equiv A 
\mathcal{C} A^T$. So, the eigenvalue problem transform as as: \begin{align} \mathcal{C}^{\prime-1} \Phi &= \Phi \Lambda \,, \\
\mathcal{C} A^{-1}\Phi &= A^{T} \Phi \Lambda\,, \end{align}
whose solution $\Phi$ isn't necessarily $A\Psi$ (unless $A$ is orthonormal). That is, affine transformation $\theta \mapsto A\theta$ does not transform PCA modes according to the same affine transformation. 

However, Eq.~\eqref{eq:geneig} transforms as: 
\begin{align} 
\mathcal{C}_{p_{A+B}}^{\prime -1}\Phi&=\mathcal{C}_{p_{A}}^{\prime -1}\Phi\Lambda \,,\\
\mathcal{C}_{p_{A+B}}^{ -1} A^{-1} \Phi&=\mathcal{C}_{p_{A}}^{ -1}A^{-1}\Phi\Lambda\,.
\end{align}
Since $A$ is non-singular, $A^{-1}\Phi=\Psi$, showing that the CPCA modes transform affinely with the parameters themselves. That is, CPCA modes represent the directions in parameter space with the most improvements in constraint, with no reference to a particular parameter basis (up to linear transforms).

We can define the relative contribution of parameter $j$ to the variance of mode $i$ by
\begin{align}\label{eq:improvement_mat}
T_{ij} = \frac{(\sqrt{\mathcal{C}_{p_{A+B}}\Psi})^2_{ij}}{\lambda_i},
\end{align}
where $\lambda_i\equiv\Lambda_{ii}$ is the $i_{\rm th}$ eigenvalue of the CPCA mode. $T_{ij}$ has the property that the sum along any row or any column sums to unity. We refer to this quantity as the improvement matrix.

To measure the number of constrained CPCA modes constrained after adding dataset $B$ we compute the number of effective modes ($N_{\rm eff}$)~\cite{Raveri:2018wln} given by
\begin{align}\label{eq:Neff}
N_{\rm eff}=N_{\rm modes}-{\rm{Tr}}(\mathcal{C}^{-1}_{p_{A}}\mathcal{C}_{p_{A+B}})=\sum\limits_{i=1}^{N_{\rm modes}}1-\lambda_i^{-1}\,,
\end{align}
where $N_{\rm modes}$ is the total number of CPCA modes. Eq.~\ref{eq:Neff} shows that larger values of $\lambda_i$ correspond to better constrained CPCA modes.

We perform the CPCA analysis and all associated calculations using the \texttt{tensiometer} package.\footnote{Available at \href{https://github.com/mraveri/tensiometer}{https://github.com/mraveri/tensiometer}}

\subsection{Kullback-Leibler Divergence}\label{Sec:KL_div}
We use the Kullback-Leibler (KL) divergence \cite{kullback} to measure the improvement in constraints of a specific parameter as a result of adding a new dataset. The KL divergence $\mathcal{D}_{\rm KL}$ quantifies the degree of similarity between two probability distributions. For posteriors $\mathcal{P}_A$ and $\mathcal{P}_{A+B}$, $\mathcal{D}_{\rm KL}$ is given by
\begin{align}
\mathcal{D}_{\rm KL}(\mathcal{P}_{A+B}\mid\mathcal{P}_A)=\int\mathcal{P}_{A+B}(\theta)\log\bigg(\frac{\mathcal{P_{A+B}}(\theta)}{\mathcal{P}_A\theta)}\bigg)d\theta\,.
\end{align}

Larger values of $\mathcal{D}_{\rm KL}(\mathcal{P}_{A+B}\mid\mathcal{P}_A)$ imply greater discrepancy between $\mathcal{P}_{A+B}$ and $\mathcal{P}_A$. Since we expect that adding to the dataset couldn't decrease the constraining power on parameters of the same model, any differences between the two posteriors can be understood to be increases in the constraining power or shifts in the posterior. In particular, when the two distributions share the same mean, $\mathcal{D}_{\rm KL}$ measures the information gain in going from $\mathcal{P}_A$ to $\mathcal{P}_{A+B}.$

\section{Datasets and Mocks} \label{Sec:datasets_mocks} 
%

\subsection{DESI Observables and Covariance}\label{Sec:desi_observables}

We generate mock observations of BAO length scales along and transverse to the line of sight with respect to $r_s$ (the sound horizon at last scattering), and the growth of structure. The length scale along the line of sight is characterized by $H(z)r_s$ and the transverse case is given by $D_A(z)/r_s$ where the angular diameter distance is
\begin{align} \label{Eq:Da}
    D_A(z)=\frac{c}{1+z}\int\limits_0^z\frac{dz'}{H(z')} \, .
\end{align}
The growth of structure is characterized by $f\sigma_8(z)$ where $f=d\ln D/dN$ is the linear growth rate and $\sigma_8(z)=\sigma_{8}D(z)$ is the amplitude of linear matter overdensity fluctuations on the $8 h^{-1}$Mpc scale at redshift $z$, where we adopt the convention that $\sigma_8\equiv\sigma_8(z=0)$, hence

\begin{align} \label{Eq:fsig8}
f\sigma_8(z)=\frac{d\ln D}{dN} \sigma_{8}D=\sigma_{8}\frac{dD}{dN}\,,
\end{align}
and $D$ is the solution to
\begin{align} \label{Eq:D}
\frac{d^2D}{dN^2}+\bigg(2+\frac{1}{H}\frac{dH}{dN}\bigg)\frac{dD}{dN}=\frac{3\Omega_{m,0}}{2a^3}\bigg(\frac{H_0}{H}\bigg)^2D \,.
\end{align}

We compute these observables for a fixed set of $\Lambda$CDM parameters and values of $\Delta \Lambda/\Lambda$. First we solve for $H(z)$ using Eqs.~\eqref{Eq:quintessenceMapping} and ~\eqref{Eq:EFTFRWequations}. We then determine the angular diameter distance using Eq.~\eqref{Eq:Da}. Finally, we integrate Eq.~\eqref{Eq:D} to determine the growth rate, $D$, and compute the RSD observable $f\sigma_8(z)$ using Eq.\eqref{Eq:fsig8}.

We calculate $H(z)r_s$, $D_A(z)/r_s$, and $f\sigma_8(z)$ in the $18$ redshift bins between $0.05$ and $1.85$ listed in Table 2.3 (baseline survey) and Table 2.5 (Bright Galaxy Survey) of \cite{Aghamousa:2017}. We also include the Lyman-$\alpha$ BAO distance measurements of $H(z)r_s$ and $D_A(z)/r_s$ using the 11 redshift bins between $1.96$ and $3.55$ listed in Table 2.7 (Ly-$\alpha$ QSO survey) of \cite{Aghamousa:2017}.

We estimate the covariance of $H(z)r_s$, $D_A(z)/r_s$, and $f\sigma_8(z)$ using the Fisher formalism assuming 14,000 sq. deg. survey and non-linear modeling up to maximum wavenumber $k_{\rm max}=0.2 \ h\rm{Mpc}^{-1}$. We verified that our forecast for the variances is consistent with Tables~2.3-2.8 of~\cite{Aghamousa:2017} for the same survey parameters. We account for correlations between $H(z)r_s,$ $D_A(z)/r_s$ and $f\sigma_8(z)$ at fixed redshift using the method outlined in  \cite{BAOcovforecast}. Correlations between $H(z)r_s$ and $D_A(z)/r_s$ are between 40\% and 42\% across all redshifts. $f\sigma_8(z)$ has a slight ($<10\%$) negative correlation with $H(z)r_s$ and $D_A(z)/r_s$.

\subsection{Baseline Dataset} \label{Subsec:dataset_description}

We perform a baseline analysis of $\Lambda$CDM and quintessence using a combination of CMB, BAO, SN, and galaxy clustering/lensing measurements. The dataset we call the ``baseline dataset" consists of the Planck 2018 measurements of CMB temperature and polarization at small (Planck 18 TTTEEE) and large angular scales (lowl+lowE) \cite{Aghanim:2018eyx, Aghanim:2019ame} and the CMB lensing potential power spectrum in the multipole range 40$\leq l \leq 400$ \cite{Aghanim:2018oex}, BAO measurements from BOSS DR12~\cite{Alam:2016hwk}, SDSS Main Galaxy Sample~\cite{Ross:2014qpa}, and 6dFGS~\cite{Beutler:2011hx}, the Pantheon Supernova (SN) sample ~\cite{Scolnic:2017caz} consisting of relative distance measurements in the redshift range $z \in [0.01, 2.26]$, and the Dark Energy Survey (DES) Year 1~\cite{Abbott:2017wau} measurement of large scale galaxy clustering, lensing, and their cross correlation ($3\times 2$). This dataset is chosen in agreement with the ``fiducial" dataset analyzed in \cite{Park:2021qnt} and we have verified that our fits to this dataset yield the same results as \cite{Park:2021qnt}.

In addition to the baseline dataset, we generate mock five-year DESI observations of the expansion rate ($H(z)r_s$), angular diameter distance ($D_A(z)/r_s$), and linear growth of structure ($f\sigma_8(z)$). As described in Sec.~\ref{Sec:desi_observables} we compute these observables for the DESI baseline survey, Bright Galaxy Survey, and Lyman-$\alpha$ forest quasar survey assuming 14000 sq. deg. of sky coverage and non-linear modelling up to $k_{\rm max}=0.2 \ h{\rm Mpc}^{-1}$. We generate one set of mocks for a $\Lambda$CDM cosmology and one set for a quintessence cosmology.

\subsection{LCDM Mocks}\label{Sec:lcdm_mocks}

To generate $\Lambda$CDM mock DESI observables that are consistent with the baseline dataset we calculate the DESI observables as described in Sec.~\ref{Sec:desi_observables} assuming $\Delta\Lambda=0$ and with $\Lambda$CDM parameters equal to their mean values obtained from fitting the baseline dataset with $\Lambda$CDM. We list these parameters in Table~\ref{tab:mean_lcdm_params}. As a check, we computed the mocks for the bestfit $\Lambda$CDM parameters and found that they agree with the mocks generated from the mean, as is expected given the $\Lambda$CDM parameter posteriors are well described by a Gaussian distribution.

\begin{table}[!htbp]
    \centering
    \begin{tabular}{ |c|c|c|c|c|c|c| } 
     \hline
    $\Omega_b h^2$ & $\Omega_c h^2$ & $\theta$ & $\tau$ & $\ln(10^{10} A_s)$ & $n_s$  & $r_s$\\ 
    \hline
     0.0224 & 0.1191 & 1.041 & 0.0582 &  3.051 & 0.9673 & 147.25 \\ 
    \hline
    \end{tabular}
   \caption{Mean $\Lambda$CDM parameters from the $\Lambda$CDM analysis of the baseline dataset.}
   \label{tab:mean_lcdm_params}
\end{table}

\begin{figure}[!b]
\includegraphics[width=\linewidth]{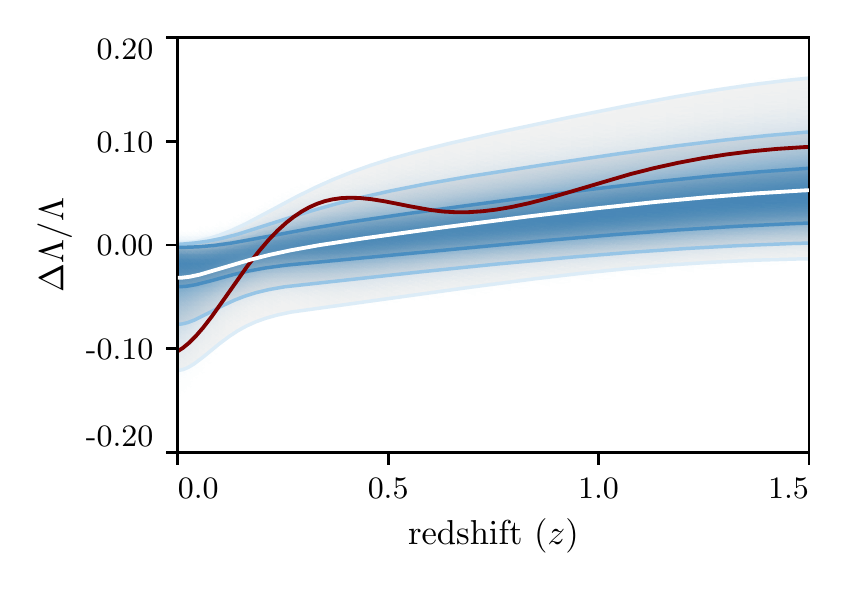}
\caption{Selected $\Delta \Lambda(z)/\Lambda$ used to generate the quintessence mock compared with the baseline dataset constraints. The white line denotes the mean from the baseline reconstruction and the shaded regions indicate the 68\%, 95\%, and 99.7\% confidence regions, respectively. The red line indicates the value of $\Delta\Lambda/\Lambda$ used to generate the quintessence mocks.}
\label{fig:delta_lambda_quint}
\end{figure}

\begin{figure*}
\includegraphics[width=\linewidth]{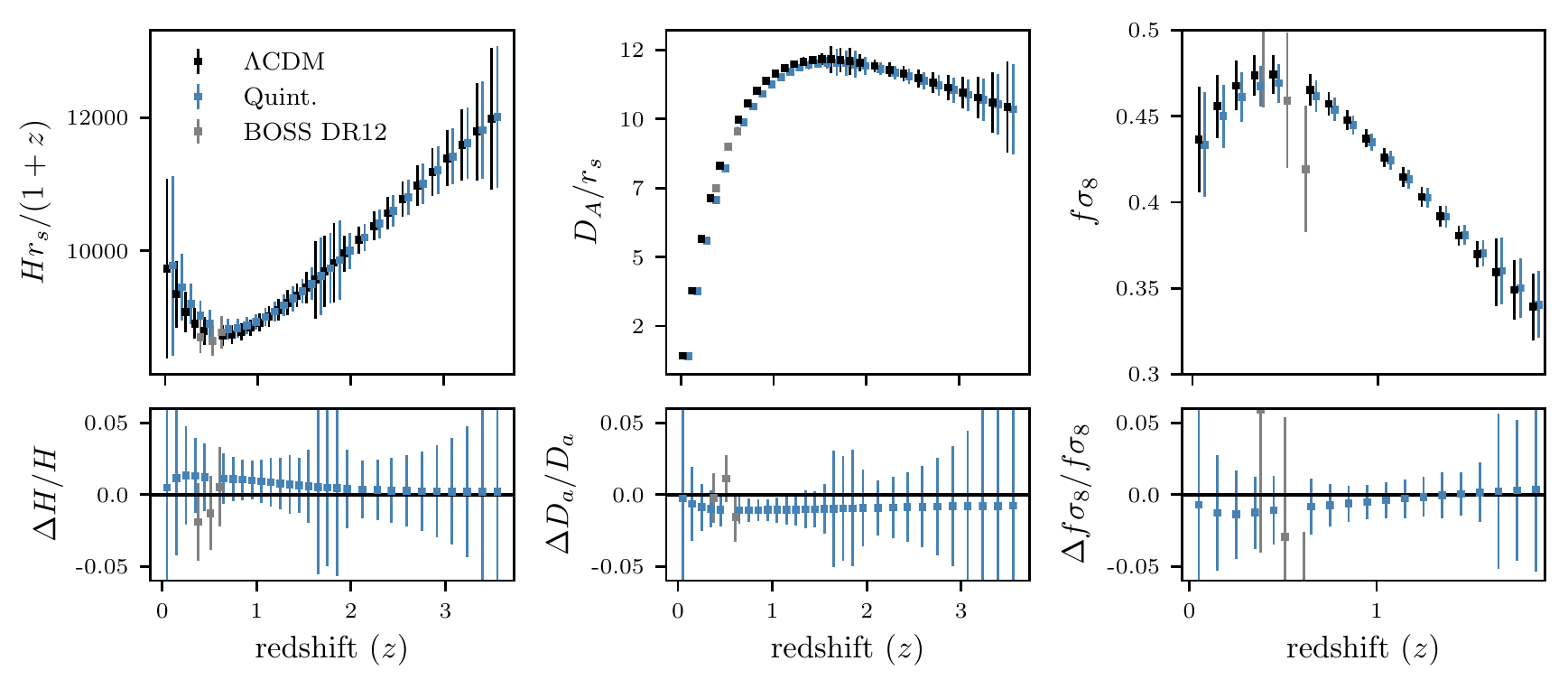}
\caption{Mock DESI observations of the expansion rate, angular diameter distance, and redshift space distortions. The black points are generated assuming $\Lambda$CDM cosmology with mean cosmological parameters derived from fitting the baseline dataset. The blue points are generated assuming a quintessence cosmology with the same $\Lambda$CDM parameters as the $\Lambda$CDM mock and a non-zero EFT function $\Delta \Lambda/\Lambda$ as detailed in Sec.~\ref{Sec:quintessence_mocks}. The horizontal offset between the black and blue points are purely for clarity as the redshift bins are the same for the two mocks. For reference we also include the BOSS DR12 measurements from \cite{Alam:2016hwk} in grey. The bottom panel shows residuals with respect to the $\Lambda$CDM cosmology.
}
\label{fig:lcdm_desi_mocks}
\end{figure*}

\subsection{Quintessence Mocks}\label{Sec:quintessence_mocks}

In order to create mock DESI observables for a quintessence cosmology we must specify the $\Lambda$CDM parameters and a non-zero $\Delta\Lambda/\Lambda$. We assume the same $\Lambda$CDM parameters as those used to generate the $\Lambda$CDM mocks. We select values of $\Delta \Lambda/\Lambda$ using the EFT reconstruction of the baseline dataset. We find all samples in the baseline dataset EFT reconstruction MCMC chain for which  $2\leq\Delta\chi^2\leq3$ with respect to $\Lambda$CDM and choose a $\Delta\Lambda/\Lambda$ from this set. Fig.~\ref{fig:delta_lambda_quint} shows our selected $\Delta\Lambda/\Lambda$ alongside the baseline constraints on $\Delta \Lambda/\Lambda.$ We compute the quintessence mock DESI observables as described in Sec.~\ref{Sec:desi_observables}. By creating our quintessence mock in this manner, we do not account for potential correlations between the EFT and $\Lambda$CDM parameters, as well as model dependent biases that can arise in the $\Lambda$CDM parameters when assuming quintessence cosmology. We discuss these subtleties in Sec.~\ref{Sec:Results}.

\subsection{Dataset Summary and Conventions}

Our final datasets consist of combinations of the baseline dataset with our mock DESI observables. We define a Fiducial $\Lambda$CDM (Quint.) dataset consisting of the baseline dataset plus all three $\Lambda$CDM (Quint.) DESI observables. In order to assess the impact of each DESI observable on our cosmological constraints, we also perform our analysis on datasets of the form baseline+$X^{\Lambda\rm{CDM}}$ and baseline+$X^{\rm{Quint.}}$ where $X \in \{Hr_s, D_A/r_s, f\sigma_8 \}$. This results in a total of nine dataset combinations that we list below:
\begin{itemize}
    \item Baseline = Planck18 TTTEEE + lowl + lowE + CMB lensing + BOSS DR12 BAO + SDSS MGS BAO + 6dFGS BAO + Pantheon SN + DES Y1 3×2
    \item Fiducial $\Lambda$CDM = Baseline+$(H+D_A+f\sigma_8)^{\Lambda \rm{CDM}}$
    \item Fiducial Quint. = Baseline+$(H+D_A+f\sigma_8)^{\rm{Quint}}$
    \item Baseline+$X^{\Lambda\rm{CDM}}; \ X \in \{H, D_A, f\sigma_8 \}$
    \item Baseline+$X^{\rm{Quint.}}; \ X \in \{H, D_A, f\sigma_8 \}$
\end{itemize}

We note that the actual DESI observations will not be completely independent of the BOSS DR12 measurements, hence including both sets of measurements in an analysis would not be appropriate. Nevertheless, since we are interested in quantifying the information gain resulting from adding DESI data, our datasets include both BOSS DR12 measurements and DESI mocks. 

 In Fig.~\ref{fig:lcdm_desi_mocks} we show our $\Lambda$CDM and quintessence mock DESI observables alongside the BOSS DR12 measurements from the full shape analysis \cite{Alam:2016hwk}. The bottom panel shows the residuals of the quintessence mock and BOSS DR12 results relative to the cosmology used to generate the $\Lambda$CDM mocks. The residuals show that DESI will measure $f\sigma_8(z)$ with significantly higher precision than BOSS, hence we expect serious gain in constraints when considering the $f\sigma_8(z)$ mocks.

\begin{figure*}
\includegraphics[width=\linewidth]{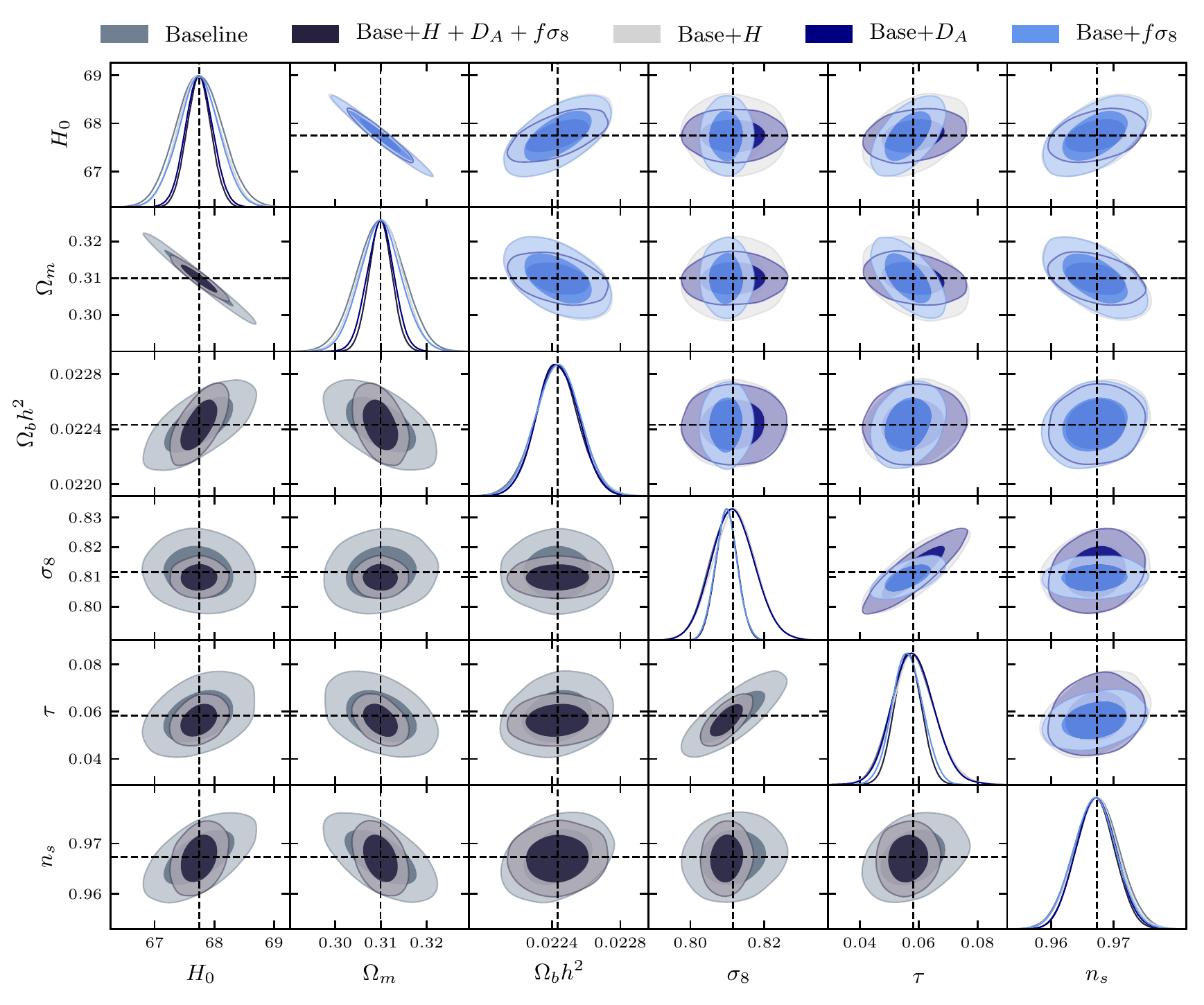}
\caption{Constraints on the six $\Lambda$CDM parameters for the $\Lambda$CDM analysis for combinations of the baseline and DESI $\Lambda$CDM datasets. The lower triangle shows the results for the baseline and fiducial DESI datasets. The upper triangle shows the results for each of the three elements of the DESI datasets. Dashed lines indicate the mean parameter values from the baseline dataset, which are also the values used to generate the DESI $\Lambda$CDM datasets.}
\label{fig:lcdm_triangle_plot}
\end{figure*}

\section{Results} \label{Sec:Results}

\subsection{LCDM improvement}\label{Sec:lcdm_improvement}

In this section we use the methods from Sec.~\ref{Sec:linear_cpca} and Sec.~\ref{Sec:KL_div} to investigate how the addition of DESI measurements of the expansion history and linear growth rate improve constraints on $\Lambda$CDM parameters.

In Fig.~\ref{fig:lcdm_triangle_plot} we show the marginalized posteriors of the six standard $\Lambda$CDM parameters for the baseline dataset, as well as combinations of the baseline dataset with the our mock DESI $\Lambda$CDM datasets. The lower triangle compares the baseline parameter constraints with the fiducial $\Lambda$CDM dataset in which we include all DESI observables. We find that including all DESI observables yields significant improvements in constraints on $H_0, \Omega_m, \sigma_8$, and $\tau.$ We also note that there is no shift in the posterior mean after adding our DESI mocks, as is to be expected since the mocks were generated from the posterior means of the baseline $\Lambda$CDM analysis.

We use the upper triangle to show the marginalized posteriors from the dataset combinations that include each DESI observable individually. Including only $H(z)r_s$ provides very little improvement in our overall parameter constraints, indicating that our baseline dataset constrains the Hubble parameter within the forecasted errors of DESI. On the other hand, adding $D_A(z)/r_s$ improves constraints on $H_0$ and $\Omega_m$ and including $f\sigma_8(z)$ improves constraints on $\sigma_8$ and $\tau.$

\begin{figure}[!t]
\includegraphics[width=0.95\linewidth]{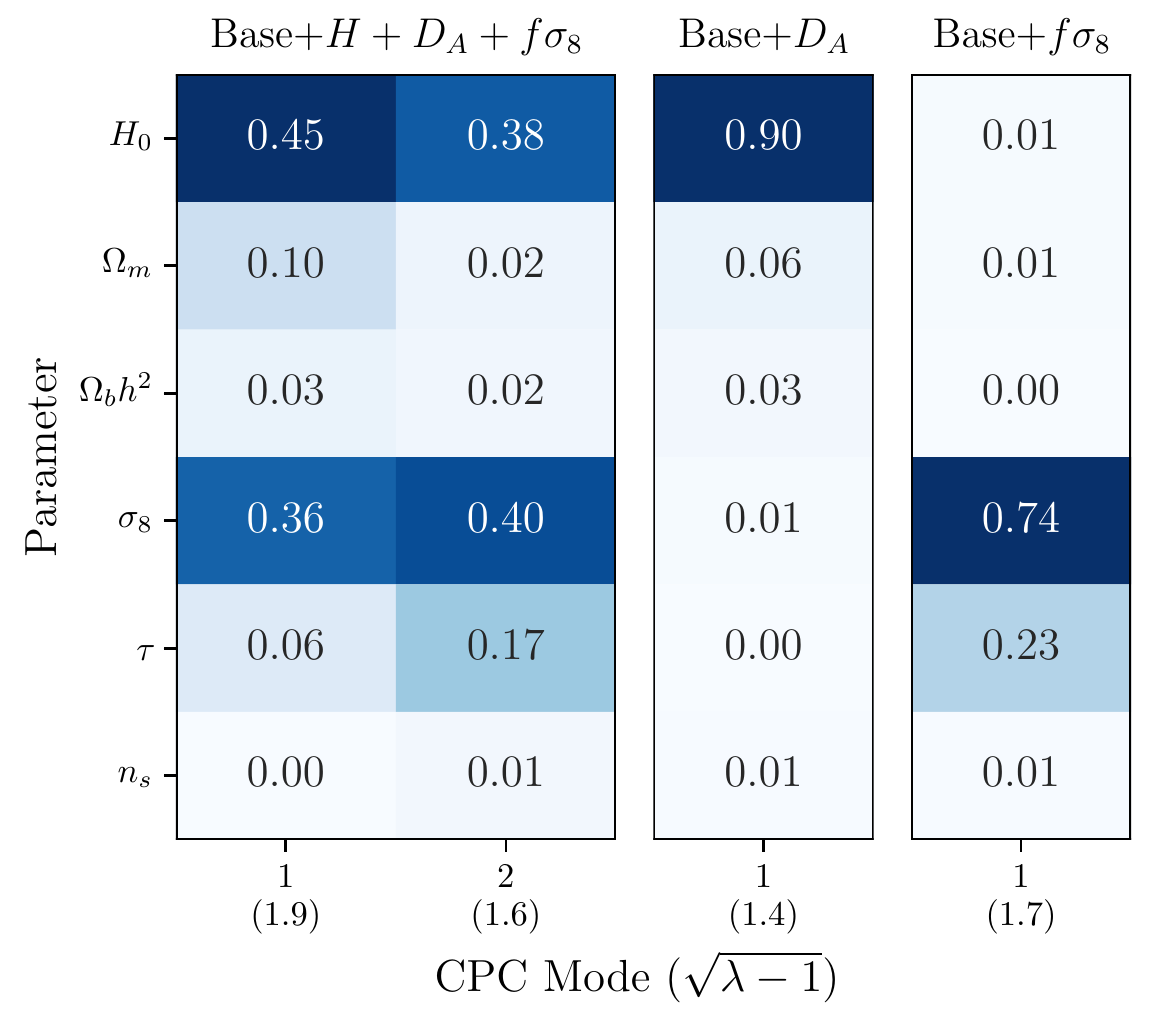}
\caption{Fractional contribution of each cosmological parameter to the variance of the labeled CPCA mode for combinations of the baseline dataset with DESI $\Lambda$CDM mocks. The value in parenthesis on the x-axis indicates the improvement of the posterior when including the DESI observations with respect to the posterior of the baseline analysis for each mode. We only show modes for which $\sqrt{\lambda-1}>1.$}
\label{fig:lcdm_frac_improve}
\end{figure}

In Fig.~\ref{fig:lcdm_frac_improve} we show the fractional improvements in the posterior after including our mock DESI observations as defined in Eq.~\eqref{eq:improvement_mat}. We only show modes for which $\sqrt{\lambda-1}>1$ which corresponds to an improvement of at least a factor of two and can be distinguished from noise induced by estimating the covariance from the MCMC samples. This cutoff choice is in agreement with the number of constrained modes, $N_{\rm eff}$, which is 0.31, 0.84, 1.12, and 1.62 for $H(z)r_s$, $D_A(z)/r_s$, $f\sigma_8(z)$, and fiducial dataset, respectively. For the fiducial $\Lambda$CDM dataset, the most improved modes are combinations of the parameters $H_0$ and $\sigma_8$. The most improved mode of the DESI angular diameter distance is dominated by $H_0$, as is expected as $D_A(z)$ is an integral of $H(z).$ This also suggests that the improved constraints on $\Omega_m$ with the addition of $D_A(z)/r_s$ in Fig.~\ref{fig:lcdm_triangle_plot} is largely a consequence of the existing degeneracy between $H_0$ and $\Omega_m.$ The best constrained mode of the $f\sigma_8(z)$ dataset is a combination of $\sigma_8$ and $\tau$.

We can further understand the relationship between these modes and the cosmological parameters by projecting the parameter samples along each CPCA mode and measuring the correlation coefficient between each projected mode and the full set of sampled and derived parameters included in the EFTCosmoMC analysis. In doing so we find that the first mode of the full DESI analysis is most correlated with $S_8$ ($r=0.99)$. The best constrained modes for the baseline+$H(z)r_s$, baseline+$D_A(z)/r_s$, and baseline+$f\sigma_8(z)$ analyses are most correlated with $H_0$ ($r=-0.9$), $hr_{\rm drag}$ ($r=-0.99$), and $\sigma_8$ ($r=0.99$), respectively.

\begin{figure}[!t]
\includegraphics[width=0.95\linewidth]{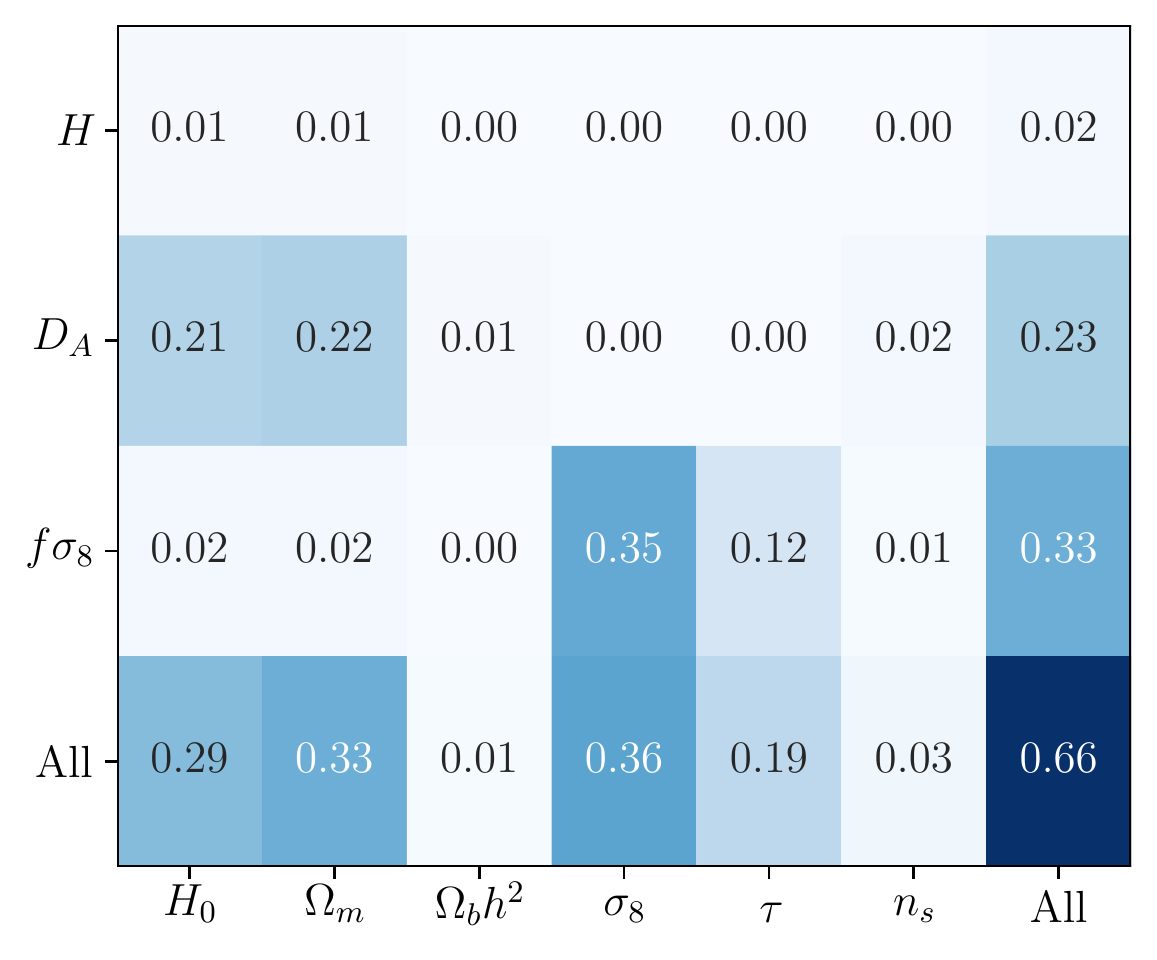}
\caption{Kullback-Leibler divergence of the posterior covariance of the baseline+DESI datasets relative to the baseline dataset for the $\Lambda$CDM analysis. $\mathcal{D}_{\rm KL}$ is shown for each parameter using their marginalized posteriors, as well as for all parameters using the full covariance of the six $\Lambda$CDM parameters after marginalizing over all nuissance parameters.}
\label{fig:lcdm_kl_div}
\end{figure}

In Fig.~\ref{fig:lcdm_kl_div} we show the KL divergence ($\mathcal{D}_{\rm KL}$) as defined in Sec.~\ref{Sec:linear_cpca} for different DESI $\Lambda$CDM datasets relative to the baseline analysis. We compute the KL divergence for each parameter using their marginalized 1D posteriors, as well as for all parameters using the full posterior of the six $\Lambda$CDM parameters marginalized over the nuisance parameters. Note that by construction the $\Lambda$CDM mocks agree with the baseline posterior means; therefore, the KL divergences measure the improvement in parameter constraints after adding DESI data.

The top row of Fig.~\ref{fig:lcdm_kl_div} shows that adding $H(z)r_s$ measurements yields essentially no improvement in our $\Lambda$CDM parameter constraints relative to the baseline dataset. From the second and third rows we again find that DESI mock observations of $D_A(z)/r_s$ significantly improves constraints on $H_0$ and $\Omega_m$ and adding mocks of $f\sigma_8(z)$ improves constraints on $\sigma_8$ and $\tau$. Including RSD measurements leads to the greatest overall improvement in $\Lambda$CDM parameter constraints, as quantified by both $\mathcal{D}_{\rm KL}$ and $N_{\rm eff}$.  The bottom row of Fig.~\ref{fig:lcdm_kl_div} shows that combining DESI measurements not only leads to improved constraints due to the sensitivity of each observable to different $\Lambda$CDM parameters, but also due to the improved constraints on specific parameters resulting from combining datasets. That is, even though $H(z)r_s$ and $f\sigma_8(z)$ provide little improvement on constraints of $H_0$ and $\Omega_m$, the fiducial dataset yields better constraints on $H_0$ and $\Omega_m$ than considering $D_A(z)/r_s$ alone.

\begin{figure*}
\includegraphics[width=\linewidth]{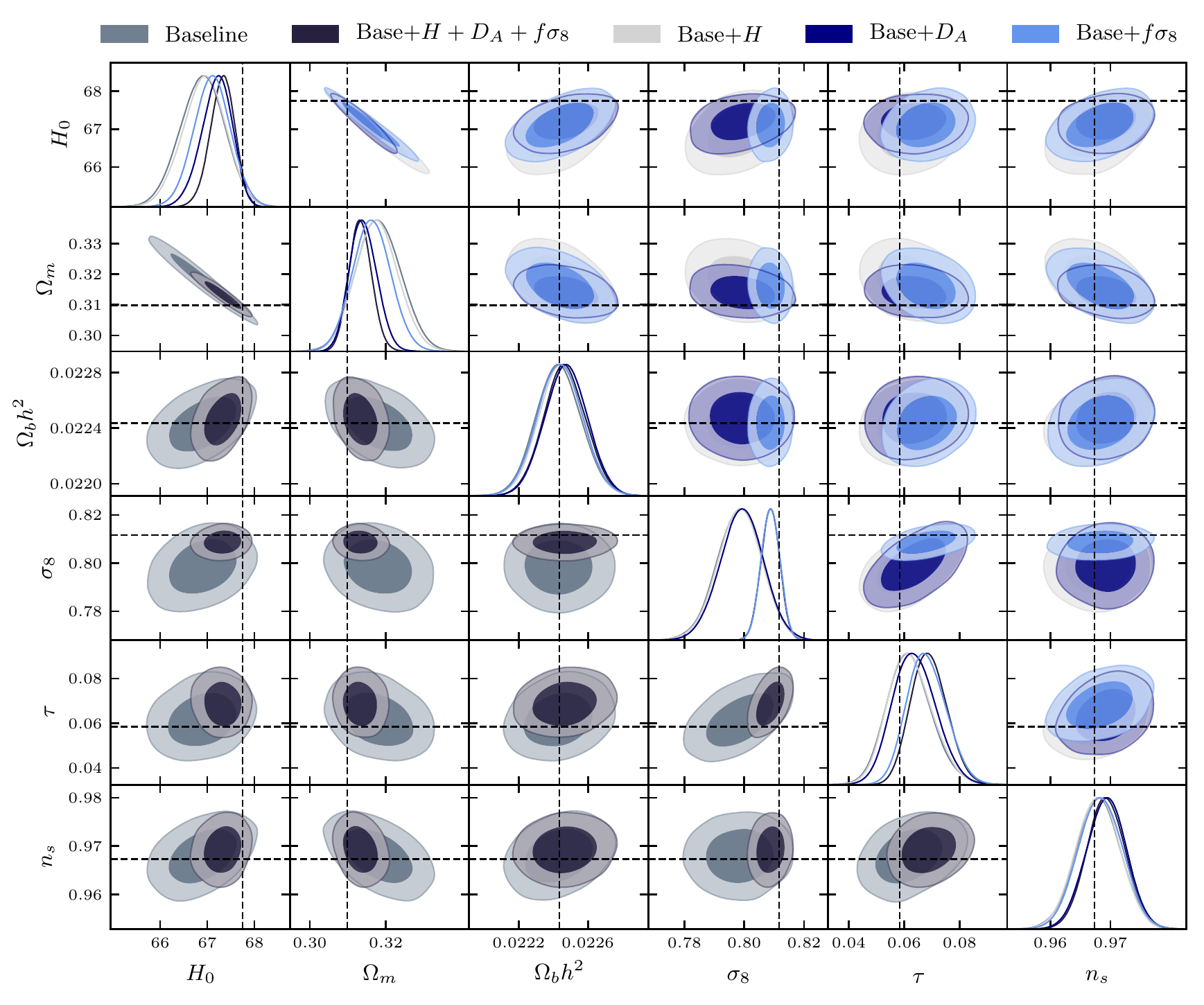}
\caption{Constraints on the six $\Lambda$CDM parameters for the EFT reconstruction of the baseline and DESI $\Lambda$CDM datasets. The lower triangle shows the results for the baseline and fiducial DESI datasets. The upper triangle shows the results for each of the three elements of the DESI datasets. Dashed lines indicate the mean parameter values from the $\Lambda$CDM analysis of the baseline dataset, which are also the values used to generate the mock DESI $\Lambda$CDM datasets.}
\label{fig:triangle_eft_lcdm_param}
\end{figure*}

Our $\Lambda$CDM analysis reveals several details about how DESI BAO distance scale and growth of structure measurements will build on the cosmological constraints from previous datasets. Firstly, there is little benefit including $H(z)r_s$ measurements in our analysis since this is already well constrained by our baseline dataset. On the contrary, DESI measurements of $D_A(z)/r_s$ and $f\sigma_8(z)$ lead to significant improvements over the baseline dataset in constraints on the $\Lambda$CDM parameters, with the most noticeable gain coming from $f\sigma_8(z)$. This is a consequence of the fact that the linear growth rate is relatively unconstrained by our baseline dataset and shows the promise of using structure formation to constrain cosmology with upcoming galaxy surveys. Finally, we find significant improvement in cosmological constraints when combining $H(z)r_s$, $D_A(z)/r_s$, and $f\sigma_8(z)$.

\subsection{Quintessence constraints}\label{Sec:quint_improvement}

We now look into how DESI observations that are consistent with a $\Lambda$CDM cosmology will impact our EFT-based reconstruction of quintessence. To this end, we analyze the quintessence reconstruction of combinations of the baseline and mock DESI datasets generated for a $\Lambda$CDM cosmology.

\begin{figure*}
\includegraphics[width=\linewidth]{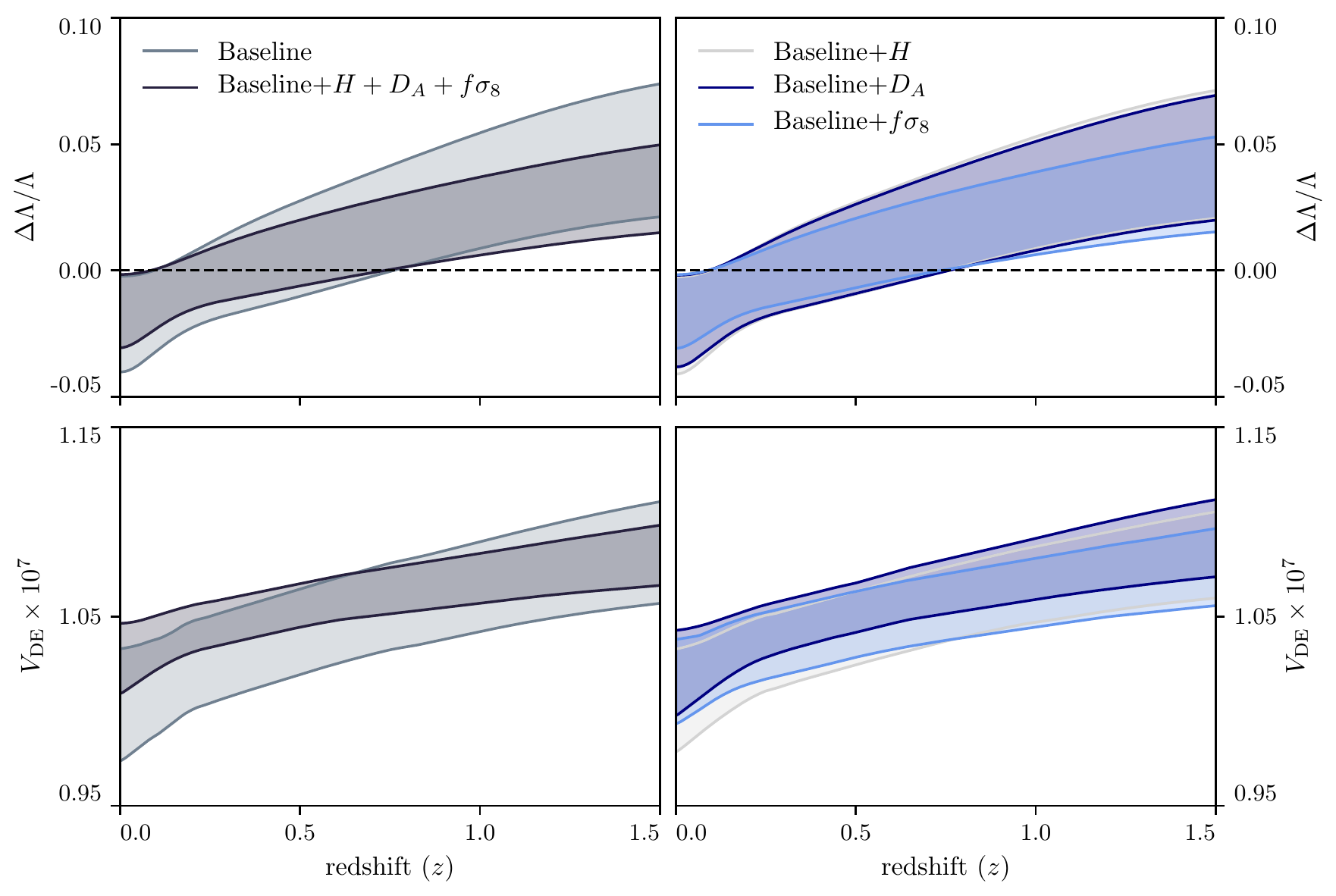}
\caption{68\% confidence interval of the reconstructed EFT function $\Delta \Lambda/\Lambda$ (top) and reconstructed potential $V_{\rm DE}(\phi(z))$ (bottom) for the baseline dataset and combinations of the baseline data with DESI $\Lambda$CDM mocks. \textit{Left}: Comparison between the baseline constraints and constraints including all DESI observables. \textit{Right}: Comparison of constraints considering each DESI observable individually.}

\label{fig:deltalambda_lcdm_desi}
\end{figure*}

First we consider how including DESI data impacts the $\Lambda$CDM parameter constraints when assuming quintessence cosmology. In Fig.~\ref{fig:triangle_eft_lcdm_param} we show the marginalized posteriors of the six $\Lambda$CDM parameters inferred from the EFT reconstruction of the baseline dataset and baseline+DESI $\Lambda$CDM datasets. Since the dashed lines represent the mean parameter values from the $\Lambda$CDM analysis of the baseline dataset, any deviations between the black lines and the baseline dataset constraints from the EFT reconstruction (light grey) are a result of the assumed cosmology. Our baseline analysis suggests that assuming quintessence cosmology leads to lower values of $H_0$ and $\sigma_8$ and higher values of $\Omega_m$ than those found when assuming $\Lambda$CDM. These findings are in agreement with existing literature showing that uncoupled quintessence cosmologies prefer lower values of $H_0$ and $\sigma_8$ than $\Lambda$CDM \cite{Banerjee:2020xcn, kunz_sigma8, doran_sigma8}. 

Including DESI mocks draws the posterior means closer to the values used to generate the mocks, and hence towards the dashed lines in Fig.~\ref{fig:triangle_eft_lcdm_param}. Since the cosmological parameters are related to the EFT function and the quintessence potential, constraints on $\Lambda$CDM parameters will be informative when analyzing the constraints on the reconstructed quintessence properties.

In Fig.~\ref{fig:deltalambda_lcdm_desi} we show the constraints on the reconstructed EFT function $\Delta \Lambda/\Lambda$ and the reconstructed quintessence potential $V(\phi(z))$ from the baseline dataset and combinations of the baseline dataset with mock DESI measurements. The left panels compare the constraints from the baseline and fiducial datasets. Including mock DESI data significantly reduces the errors on $\Delta \Lambda/\Lambda$ and brings the mean value of the EFT function closer to zero, the $\Lambda$CDM limit. The bottom panel shows that the inclusion of DESI data consistent with a $\Lambda$CDM cosmology significantly improves constraints on the potential. The reconstructed potential of the fiducial analysis is flatter than that of the baseline analysis as is consistent with the $\Lambda$CDM limit of quintessence. 

The right panels of Fig.~\ref{fig:deltalambda_lcdm_desi} shows the constraints on $\Delta \Lambda/\Lambda$ and $V(\phi(z))$ after considering each DESI observable individually. Whereas measurements of $H(z)r_s$ and $D_A(z)/r_s$ provide little improvement on the constraints on the EFT function, including $f\sigma_8(z)$ significantly constrains $\Delta \Lambda/\Lambda$. On the other hand, constraints on $V_{\rm DE}$ come from both angular diameter distance and growth rate measurements. Furthermore, the addition of $D_A(z)/r_s$ measurements increases the amplitude of the quintessence potential. This is a consequence of the fact that the posterior of $H_0$ increases after adding our $D_A(z)/r_s$ measurements, and the amplitude of the potential is sensitive to $H_0.$ To make quantitative statements about how DESI measurements will constrain the EFT function and reconstructed potential we perform a CPCA analysis of these quantities.

\begin{figure*}
\includegraphics[width=0.9\linewidth]{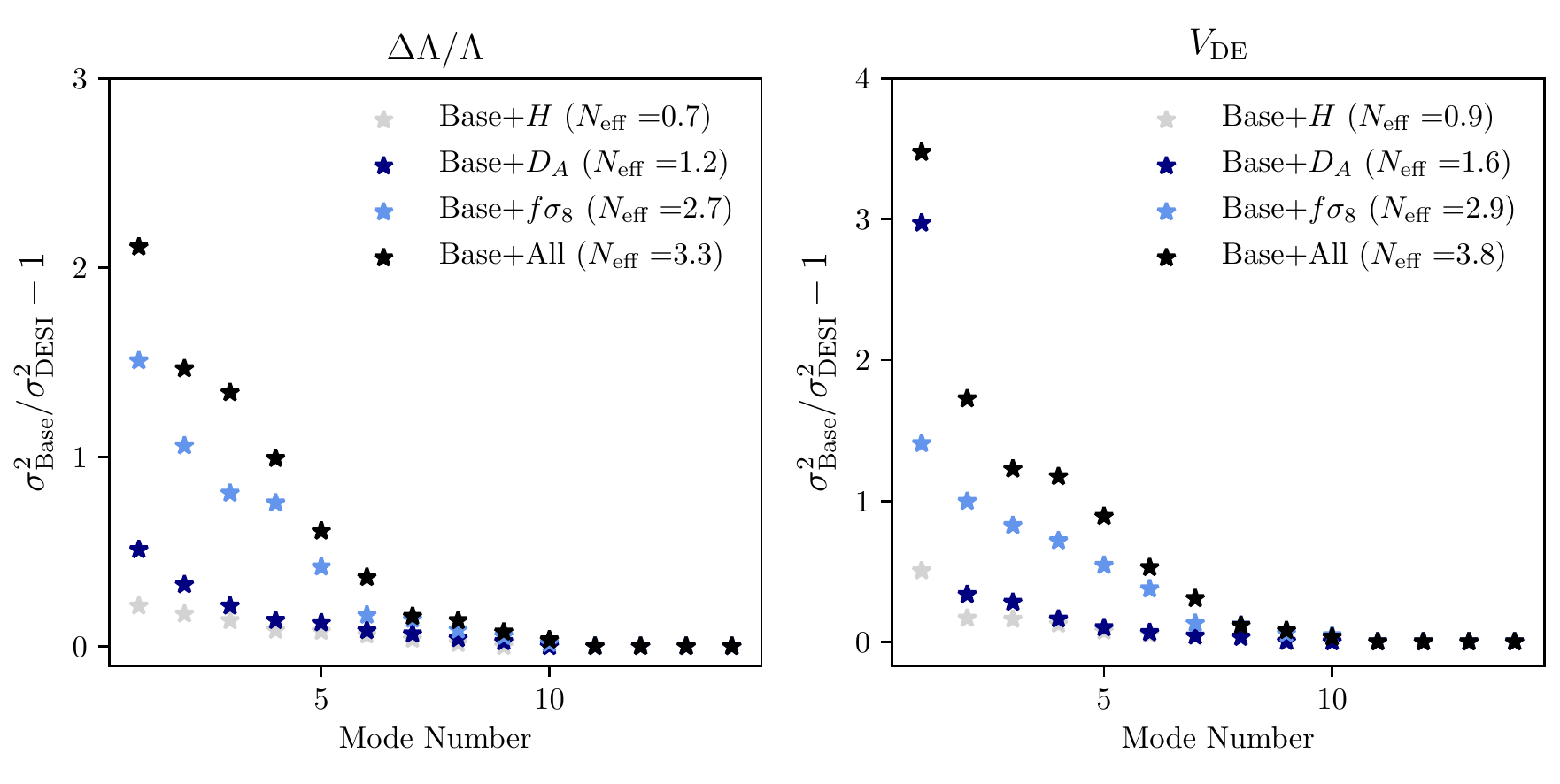}
\caption{Sorted CPCA eigenvalues computed from the posterior covariance of the DESI $\Lambda$CDM analyses relative to that of the baseline analysis for the EFT function $\Delta\Lambda/\Lambda$ (left) and reconstructed quintessence potential $V(\phi)(z)$ (right). The effective number of constrained modes $N_{\rm eff}$ for each dataset and reconstructed function is shown in the legend.}
\label{fig:lambda_cpca_eig_improvement}
\end{figure*}

In Fig.~\ref{fig:lambda_cpca_eig_improvement} we plot the sorted CPCA eigenvalues, as defined in Sec.~\ref{Sec:linear_cpca} for the EFT function (left) and reconstructed quintessence potential (right) for different combinations of the DESI $\Lambda$CDM analyses relative to the baseline analysis. The CPCA analyses of the EFT function $\Delta\Lambda/\Lambda$ and reconstructed potential $V(\phi(z))$ are performed using the posterior covariances as estimated from the MCMC samples of the 14 $\Delta\Lambda/\Lambda$ and $V(\phi(z))$ values, respectively. The CPCA analysis of $\Delta \Lambda/\Lambda$ again shows that including $H(z)r_s$ measurements provides essentially no improvement on our constraints of $\Delta \Lambda/\Lambda$ relative to the baseline dataset. Adding measurements of the angular diameter distance results in slightly better constraints on $\Delta \Lambda/\Lambda$; however, the most significant improvement results from adding measurements of the growth rate. The extent of the improvement can be quantified via the number of constrained modes  $(N_{\rm eff})$ relative to the baseline dataset as introduced in Sec.~\ref{Sec:linear_cpca}. Including all elements in the mock DESI dataset increases the number of constrained modes by $3.3$, where the majority of the improvement comes from the growth rate measurements which constrain 2.7 modes on their own.

In  the right panel of Fig.~\ref{fig:lambda_cpca_eig_improvement} we show the eigenvalues of the CPCA analysis of the reconstructed quintessence potential. Unlike the CPCA analysis of $\Delta \Lambda/\Lambda$ in which $f\sigma_8(z)$ was the most significant source of improvement across all modes, the first mode of the reconstructed potential is most improved by the addition of the angular diameter distance measurements. Similarly, there is an uptick in the first eigenvalue of the Hubble parameter analysis; therefore, the $H(z)r_s$ measurements inform the constrains on one mode of the potential. The $N_{\rm eff}$ values for the reconstructed potential are also larger than those of the EFT function for all dataset combinations, indicating that our DESI mocks better constrain the potential than the EFT function. In order to understand the source of the differences between constraints on $\Delta\Lambda/\Lambda$ and $V(\phi(z))$, it is useful to plot the CPCA modes.

\begin{figure*}
\includegraphics[width=\linewidth]{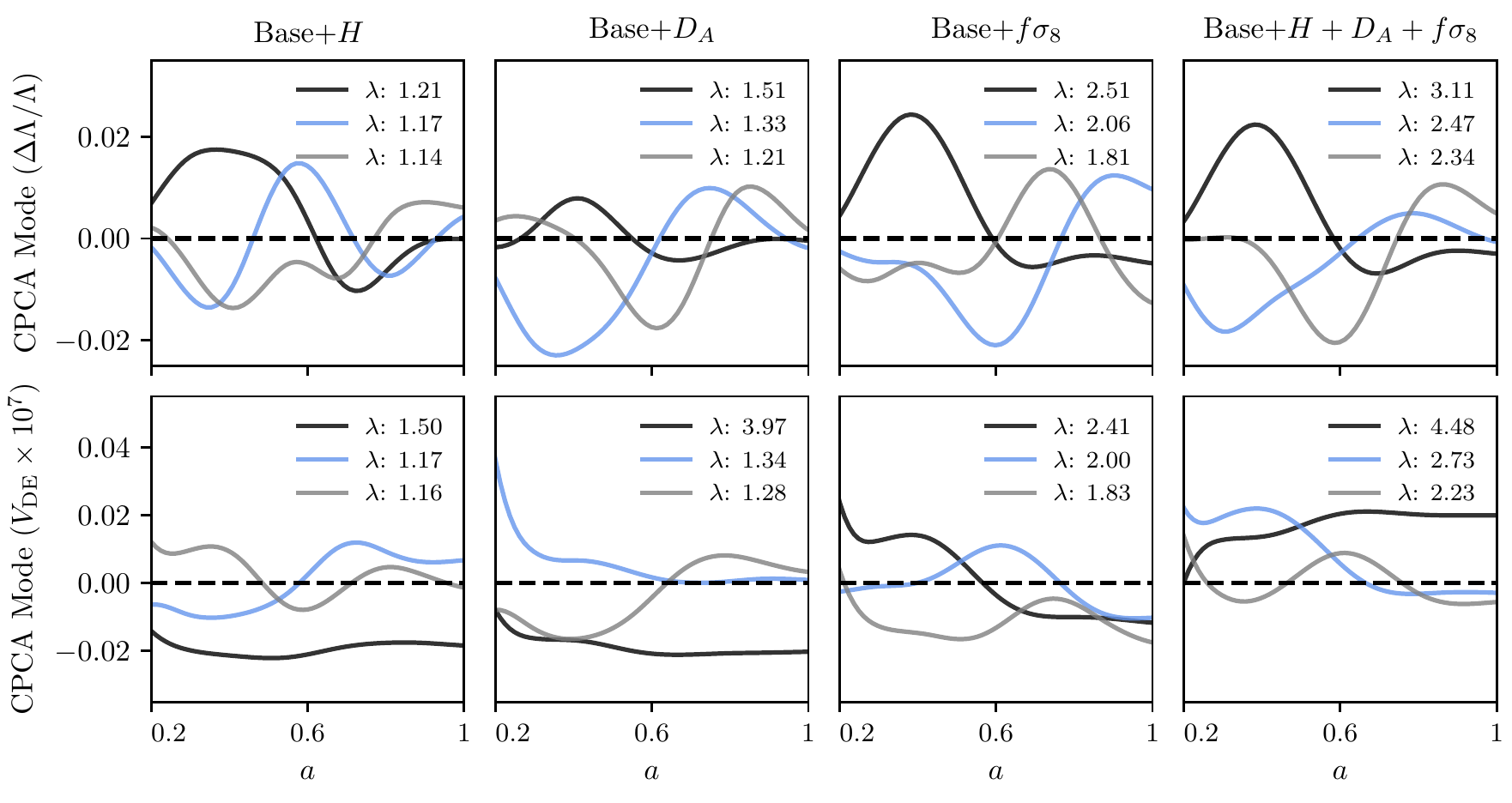}
\caption{\textit{Top}: the three best constrained CPCA modes of the reconstructed EFT function for combinations of the baseline dataset with DESI $\Lambda$CDM mocks. \textit{Bottom}: the three best constrained CPCA modes of the quintessence potential for combinations of the DESI $\Lambda$CDM datasets. The corresponding CPCA eigenvalues are shown in the legend.
}
\label{fig:cpca_mode}
\end{figure*}

Fig.~\ref{fig:cpca_mode} shows the three most significant CPCA modes for the reconstructed EFT function (top) and the reconstructed quintessence potential (bottom) using the DESI $\Lambda$CDM datasets. From the top panels we see that the best constrained mode of the fiducial DESI dataset is nearly identical to the best constrained mode of the $f\sigma_8(z)$ dataset. Furthermore, the second best constrained mode of the growth rate analysis is similar to the third best constrained mode of the combined analysis. This supports our conclusion that the improvement in constraints on the EFT function in the combined analysis of the DESI $\Lambda$CDM mocks stems predominately from the $f\sigma_8(z)$ measurements.

The bottom panels of Fig.~\ref{fig:cpca_mode} shows the three best constrained CPCA modes of the reconstructed potential. The best constrained mode is flat for all dataset combinations except the $f\sigma_8(z)$ measurements. This reflects the fact that the amplitude of the potential is related to $H_0$ and $\Omega_m$ which is best constrained by the inclusion of the BAO measurements. On the other hand, $f\sigma_8(z)$ is relatively insensitive to the amplitude of the potential, and hence the best constrained mode from the growth rate measurements is not constant across scale factors.

To verify that the constant mode is a result of improved constraints on $\Omega_m$ and $H_0$, we compute the correlation coefficient between the CPCA modes and the $\Lambda$CDM parameters and find that the best constrained modes of the $H(z)r_s$, $D_A(z)/r_s,$ and fiducial analyses are correlated with $H_0$ and $\Omega_m$ by more than 80\%. For the $f\sigma_8(z)$ analysis, all correlations between the CPCA modes and $H_0$ and $\Omega_m$ are below 30\%. Moreover, the correlations between the top four constrained modes of the $f\sigma_8(z)$ dataset and the $\Lambda$CDM parameters are all below 50\%, hence the constraints on the potential resulting from the growth rate measurements cannot be accounted for by improvements in the $\Lambda$CDM parameter constraints. Finally, we note that the best constrained mode of the $f\sigma_8(z)$ analysis is the same as the second best constrained mode in the combined analysis. This is consistent with the ordering of the eigenvalues in Fig.~\ref{fig:lambda_cpca_eig_improvement}.

Having performed the CPCA analyses of the reconstructed EFT function and the potential, we return to the question of why the $\Lambda$CDM angular diameter distance measurements do not improve constraints on the EFT function, $\Delta \Lambda/\Lambda$? From Eq.~\ref{Eq:quintessenceMapping} we expect $\Lambda\approx -V(\phi(a))$ for a potential energy dominated field; therefore, we expect comparable constraints on the EFT function and the reconstructed potential. However, recall that the EFT function we reconstruct is computed relative to its $\Lambda$CDM value as defined in Eq.~\ref{Eq:EFT_function}. In effect, the rescaled EFT function has an explicit, non-trivial dependence on $H_0$ and $\Omega_m$. The relative EFT function $\Delta \Lambda/\Lambda$ is then largely insensitive to the $\Lambda$CDM BAO measurements because their impact can be fully absorbed in the constraints on $\Omega_m$ and $H_0$. This also  explains the high correlation between the first CPCA mode of the reconstructed potential and $H_0$ and $\Omega_m$ when including BAO measurements. To check this, we performed a CPCA analysis of the full EFT function $\Lambda$, which has an explicit dependence on $\Delta \Lambda/\Lambda, \Omega_m, $ and $H_0$, and found the results were consistent with those of the quintessence potential. These conclusions no longer apply when analyzing a mock for which constraints on the potential cannot be fully accounted for by shifts in the $\Lambda$CDM parameters as is described in Sec.~\ref{Sec:quint_detection}.

\begin{figure}[!t]
\includegraphics[width=0.9\linewidth]{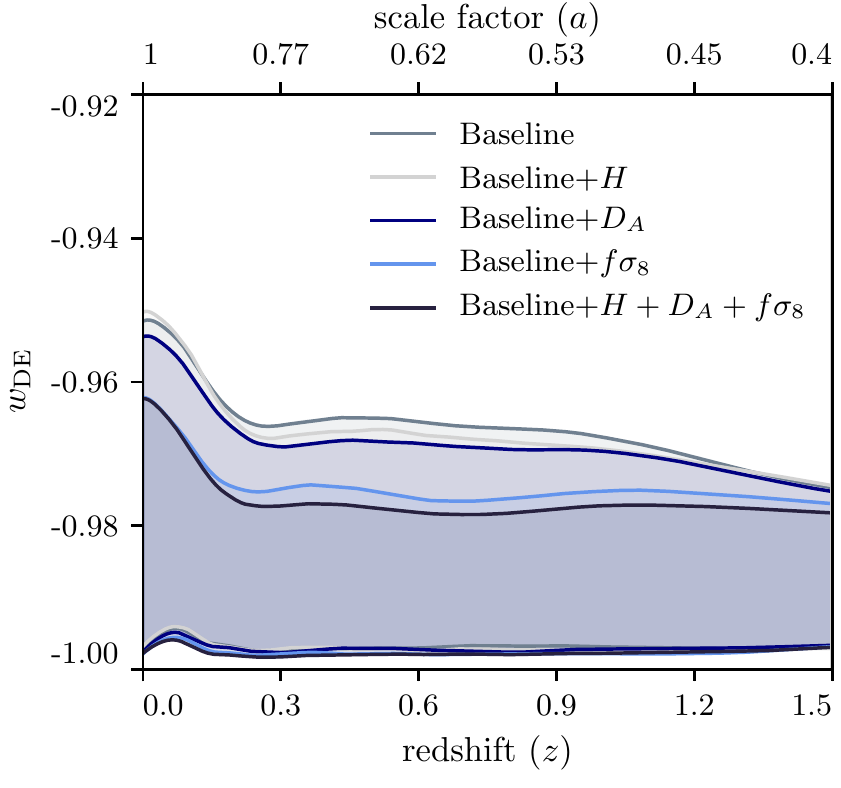}
\caption{68\% C.L. on the dark energy equation of state $w_{\rm DE}(z)$ as a function of redshift from the baseline dataset and combinations of the baseline data with DESI $\Lambda$CDM mocks.}
\label{fig:wde_improvement}
\end{figure}

\begin{figure*}
\includegraphics[width=\linewidth]{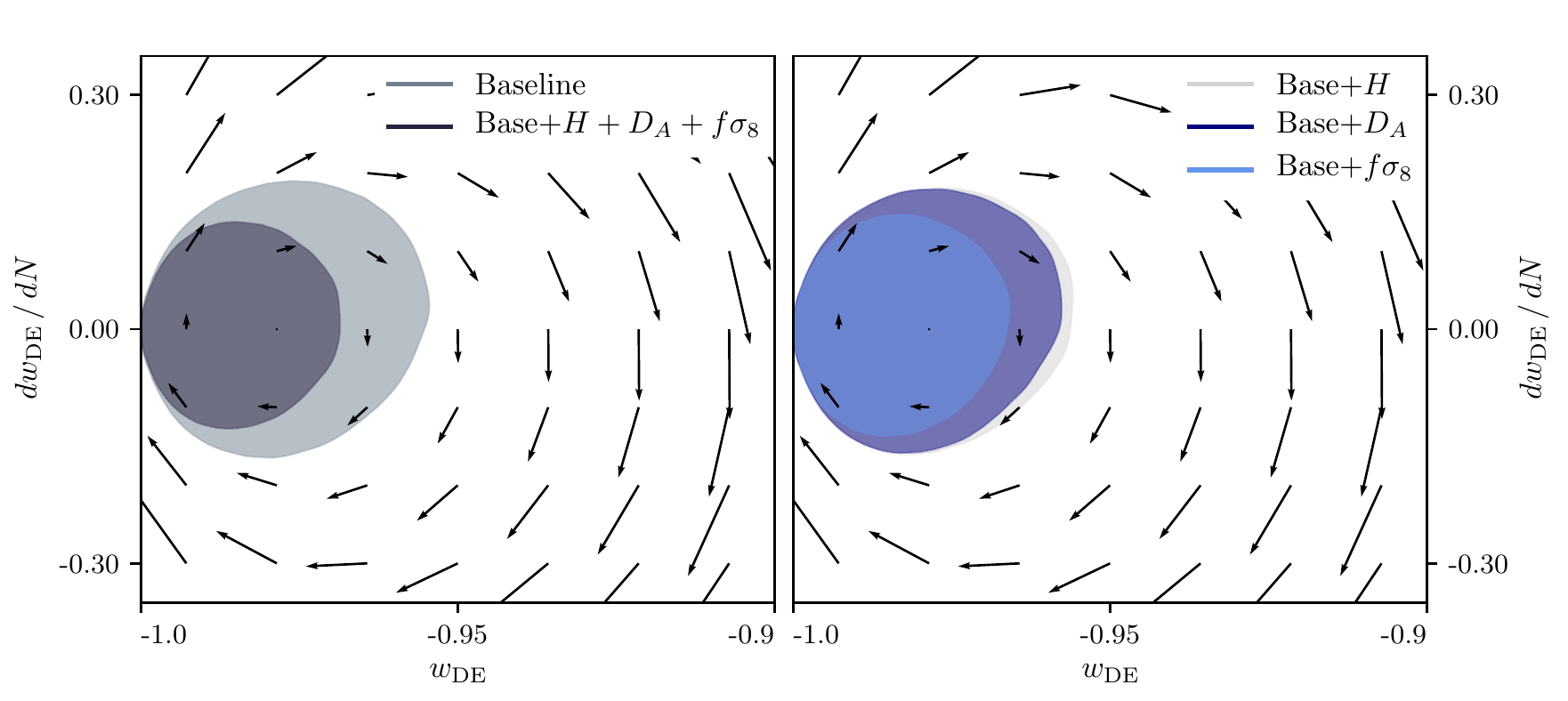}
\caption{Reconstructed trajectories in the of $w_{\rm DE}, dw_{\rm DE}/dN$ phase space at the 68\% C.L.for different DESI dataset combinations. The arrows show the average velocity at a given point in phase space for the baseline dataset reconstruction.}
\label{fig:w_DE_desi_phasespace}
\end{figure*}

In Fig.~\ref{fig:wde_improvement} we show the 68\% C.L. of the reconstructed dark energy equation for the baseline and DESI $\Lambda$CDM datasets we consider. We find that the addition of DESI observations consistent with $\Lambda$CDM significantly tightens constraints on $w_{\rm DE}$, with the constraining power coming almost entirely from the $f\sigma_8(z)$ mocks. Our finding that $w_{\rm DE}$ is particularly sensitive to growth of structure measurements from DESI is in agreement with the results of \cite{Perenon} and is indicative of the power of using structure information to constrain dynamical dark energy models, even if such models do not have significant perturbations.

Using the results of the EFT reconstruction, we can reconstruct the trajectories in the $(w_{\rm DE}, dw_{\rm DE}/dN)$ phase space. The $(w_{\rm DE}, dw_{\rm DE}/dN)$ phase space is useful for distinguishing between freezing and thawing models of quintessence \cite{Caldwell:2005ai}. \citet{Park:2021qnt} showed that the $(w_{\rm DE}, dw_{\rm DE}/dN)$ trajectories for the baseline dataset reconstruction follow neither freezing nor thawing scenarios, but instead oscillate about the boundary at $w_{\rm DE}=-1$. In Fig.~\ref{fig:w_DE_desi_phasespace} we show the phase space distribution of trajectories in the $w_{\rm DE}-dw_{\rm DE}/dN$ plane for elements of the mock $\Lambda$CDM DESI dataset. The arrows indicate the average velocity at each point in the phase space for the baseline dataset and the contours show the 68\% confidence intervals. We find that DESI will significantly improve constraints on the distribution of trajectories in the $(w_{\rm DE}, dw_{\rm DE}/dN)$ phase space. This improvement comes almost entirely from the growth rate measurements, which provide our tightest constraints on both the dark energy equation of state and its time dependence. Improved constraints on the $(w_{\rm DE}, dw_{\rm DE}/dN)$ phase space will restrict the extent to which observationally viable models of quintessence can possess dynamical behavior, and hence deviate from a $\Lambda$CDM cosmology.

Finally, we investigate how DESI will improve constraints on the Swampland Conjectures for quintessence models. We found that adding DESI data consistent with a $\Lambda$CDM cosmology provides little improvement on constraints of $V_\phi/V$ and $\nabla^2_\phi V/V$, hence it is unlikely that we will be able to place tighter bounds on $|V_\phi/V|$ and $-\nabla^2_\phi V/V$ with DESI using the EFT reconstruction. This is to be expected given that \cite{Park:2021qnt} found no evidence for lower bounds on $|\nabla_\phi V|/V$ and $-\nabla^2_\phi V/V$ and the mock data used in this analysis violates Swampland as it was generated for a $\Lambda$CDM cosmology.

\begin{figure}[!t]
\includegraphics[width=0.99\linewidth]{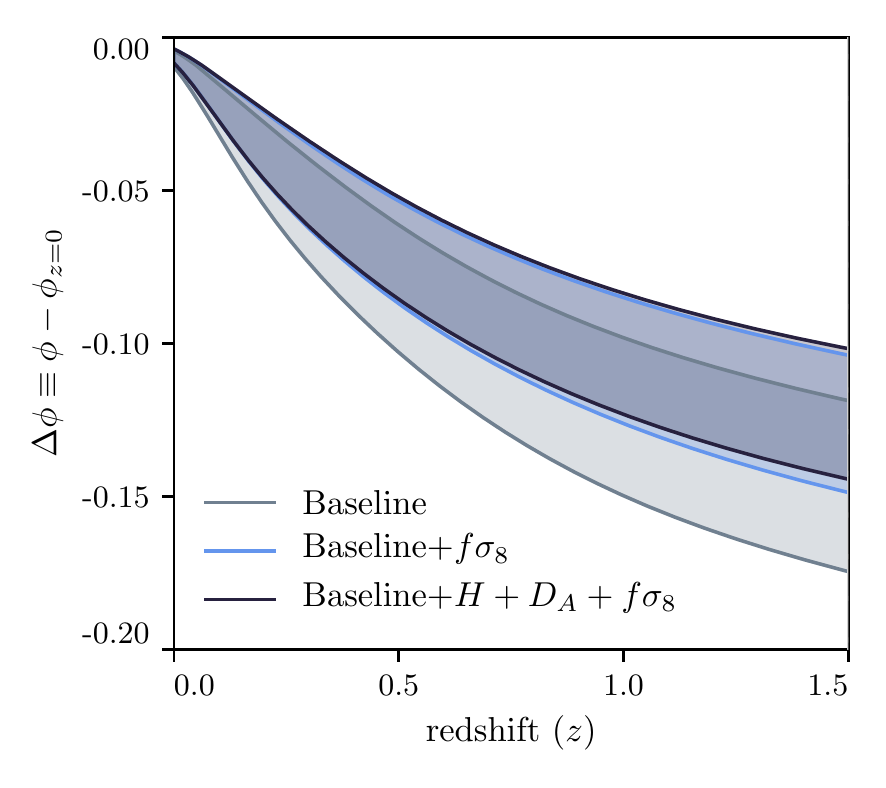}
\caption{68\% C.L. of the scalar field excursion from the baseline dataset and combinations of the baseline data with DESI $\Lambda$CDM mocks.} 
\label{fig:field_exc}
\end{figure}

\begin{figure*}
\includegraphics[width=\linewidth]{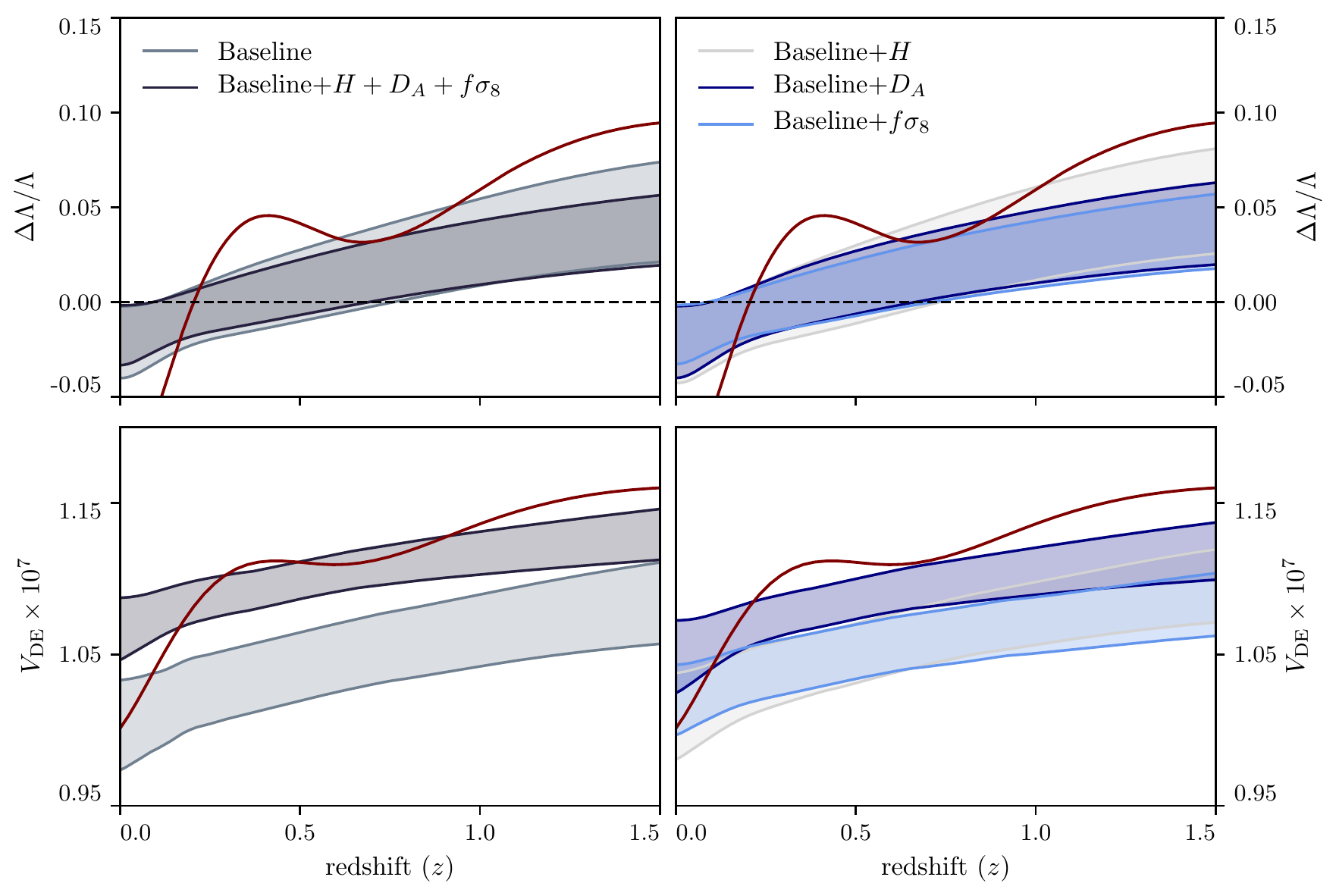}
\caption{68\% confidence interval of the reconstructed EFT function $\Delta \Lambda/\Lambda$ (top) and reconstructed potential $V_{\rm DE}(\phi(z))$ (bottom) for the baseline dataset and combinations of the baseline data with DESI quintessence mocks. \textit{Left}: Comparison between the baseline constraints and constraints including all DESI observables. \textit{Right}: Comparison of constraints considering each DESI observable individually. The red line indicates the values of $\Delta\Lambda/\Lambda$ and $V(\phi(z))$ used to generate the quintessence mock  datasets.}
\label{fig:fig6_desi_quint}
\end{figure*}

In addition to placing boundaries on the derivatives of the scalar field potential, the Swampland Conjectures impose restrictions on the magnitude of the scalar field excursion $|\Delta \phi|$, where $\Delta\phi(z)\equiv \phi(z)-\phi(z=0)$. In Fig.~\ref{fig:field_exc} we show the 68\% C.L. on the scalar field excursion as a function of redshift for the baseline dataset, as well as combinations of the baseline dataset with DESI $\Lambda$CDM datasets. We only show the individual dataset results for the $f\sigma_8(z)$ analysis since including $H(z)r_s$ or $D_A(z)/r_s$ measurements do not improve our constraints on $\Delta \phi(z)$. If DESI observes a Universe consistent with $\Lambda$CDM cosmology, then the method used in this work will constrain $|\Delta\phi|\lesssim0.12$ and $0.15$ for $z<1.5$ at the 68\% and 95\% C.L., respectively.

Ultimately, the EFT reconstruction of the $\Lambda$CDM datasets suggests that a DESI observation consistent with $\Lambda$CDM cosmology will significantly constrain the space of viable quintessence models. Our results agree with the findings of \cite{flat_phiCDM} which used Bayesian model selection techniques to address the extent to which DESI observations from a $\Lambda$CDM cosmology can rule out various quintessence and phantom dark energy models.

\subsection{Quintessence detection}\label{Sec:quint_detection}

Having analyzed how DESI will impact constraints on quintessence if it observes a Universe consistent with $\Lambda$CDM, we now explore how mock DESI observations generated from a quintessence cosmology will impact the results of the EFT reconstruction. Since there is a priori no well motivated choice for the quintessence cosmology from which we compute the mocks, we emphasize that the results in this section are heavily dependent on the choice of the quintessence cosmology. Furthermore, the choice of the quintessence mock is complicated by the fact that the EFT reconstruction and $\Lambda$CDM analyses prefer different cosmological parameters as discussed in Sec.~\ref{Sec:quint_improvement} and the reconstructed EFT function can be degenerate with these parameters. This can lead to mocks that are in tension with the baseline dataset, and hence unrealistically tight parameter constraints. In light of these concerns, we use the quintessence mocks analyzed in this section to study whether or not we can identify any interesting features present in the mocks and leave the discussion of overall improvement in parameter constraints to Sec.~\ref{Sec:lcdm_improvement} and ~\ref{Sec:quint_improvement}.

\begin{figure*}
\includegraphics[width=0.9\linewidth]{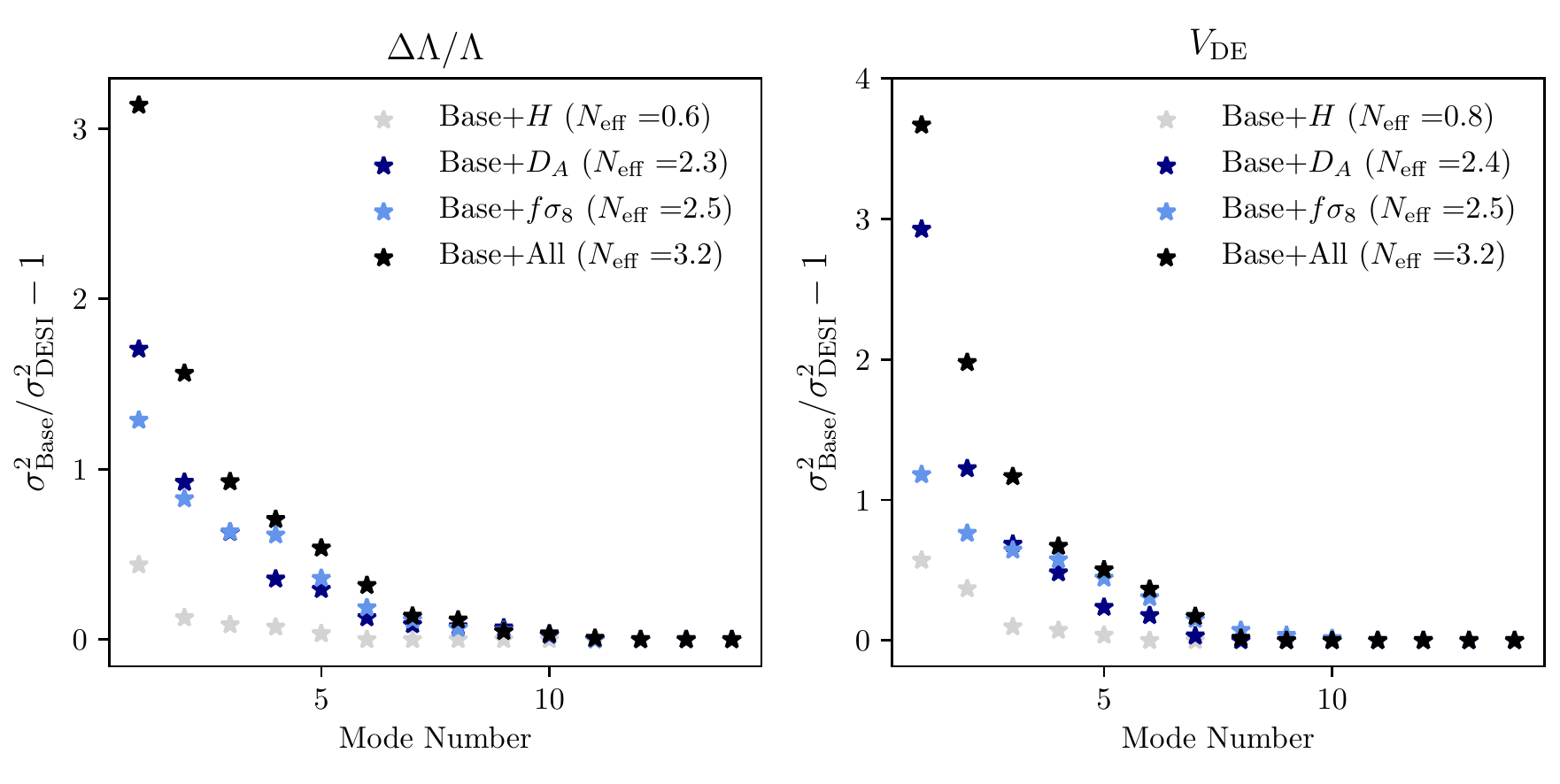}
\caption{Sorted CPCA eigenvalues computed from the posterior covariance of the DESI quintessence analyses relative to that of the baseline analysis for the EFT function $\Delta\Lambda/\Lambda$ (left) and reconstructed quintessence potential $V(\phi)(z)$ (right). The effective number of constrained modes $N_{\rm eff}$ for each dataset is shown in the legend.}
\label{fig:Lambda_cpca_eig_improvement_quint}
\end{figure*}

In Fig.~\ref{fig:fig6_desi_quint} we show the constraints on the reconstructed EFT function $\Delta \Lambda/\Lambda$ and the reconstructed quintessence potential $V(\phi(z))$ from the baseline dataset and combinations of the baseline dataset with DESI observables generated for a quintessence cosmology. The red lines are used to indicate the value of $\Delta \Lambda/\Lambda$ and $V(\phi(z))$ used to generate the DESI mocks. Although the constraints on $\Delta \Lambda/\Lambda$ tighten after adding the DESI datasets, the posterior of the reconstructed potential does not shift towards the value used to generate the DESI mocks. Nevertheless, the reconstructed EFT function does not shift as close to zero as it did for the $\Lambda$CDM mocks. The angular diameter distance measurements from the quintessence mock constrain the reconstructed EFT function significantly more than the $\Lambda$CDM mock datasets.

In the bottom panel of  Fig.~\ref{fig:fig6_desi_quint} we show the constraints on the quintessence potential. We immediately see that the amplitude of the quintessence mock is much larger than of the $\Lambda$CDM mock. This is due to the fact that the value of $H_0$ used to generate the quintessence mock is the mean from the baseline $\Lambda$CDM constraints, and hence larger than the $H_0$ value preferred by the EFT reconstruction of the baseline dataset. Although this choice places our quintessence mock in tension with the baseline constraints, after including all DESI measurements we are able to recover the amplitude of the potential and significantly reduce the error on the reconstructed potential. We note that some of this reduction in error is sourced by the underlying tension between the baseline and mock DESI datasets.

In Fig.~\ref{fig:Lambda_cpca_eig_improvement_quint} we plot the sorted CPCA eigenvalues, as defined in Sec.~\ref{Sec:linear_cpca}, for the EFT function (left) and reconstructed quintessence potential (right) for different combinations of the DESI quintessence analyses relative to the baseline analysis. The CPCA analyses of the EFT function $\Delta\Lambda/\Lambda$ and reconstructed potential $V(\phi(z))$ are performed using the posterior covariances as estimated from the MCMC samples of the 14 $\Delta\Lambda/\Lambda$ and $V(\phi(z))$ values, respectively. The results of the left panel suggest that the angular diameter distance measurements are significantly more constraining for the EFT function of the quintessence mocks than they were for the $\Lambda$CDM mocks. The quintessence $D_A(z)/r_s$ measurements constrain approximately the same number of modes of the EFT functions as the quintessence $f\sigma_8(z)$ measurements.

The right panel of Fig.~\ref{fig:Lambda_cpca_eig_improvement_quint} shows the CPCA eigenvalues of the reconstructed quintessence potential for the DESI quintessence mocks. As was the case with the $\Lambda$CDM datasets, we find that the first mode of the potential is predominately constrained by the angular diameter distance measurements; however, unlike in the analysis of the $\Lambda$CDM mocks, the angular diameter distance measurements for the quintessence mocks provide significant constraining power beyond just the first mode. We can determine the source of these constraints by plotting the CPCA modes.

\begin{figure*}
\includegraphics[width=\linewidth]{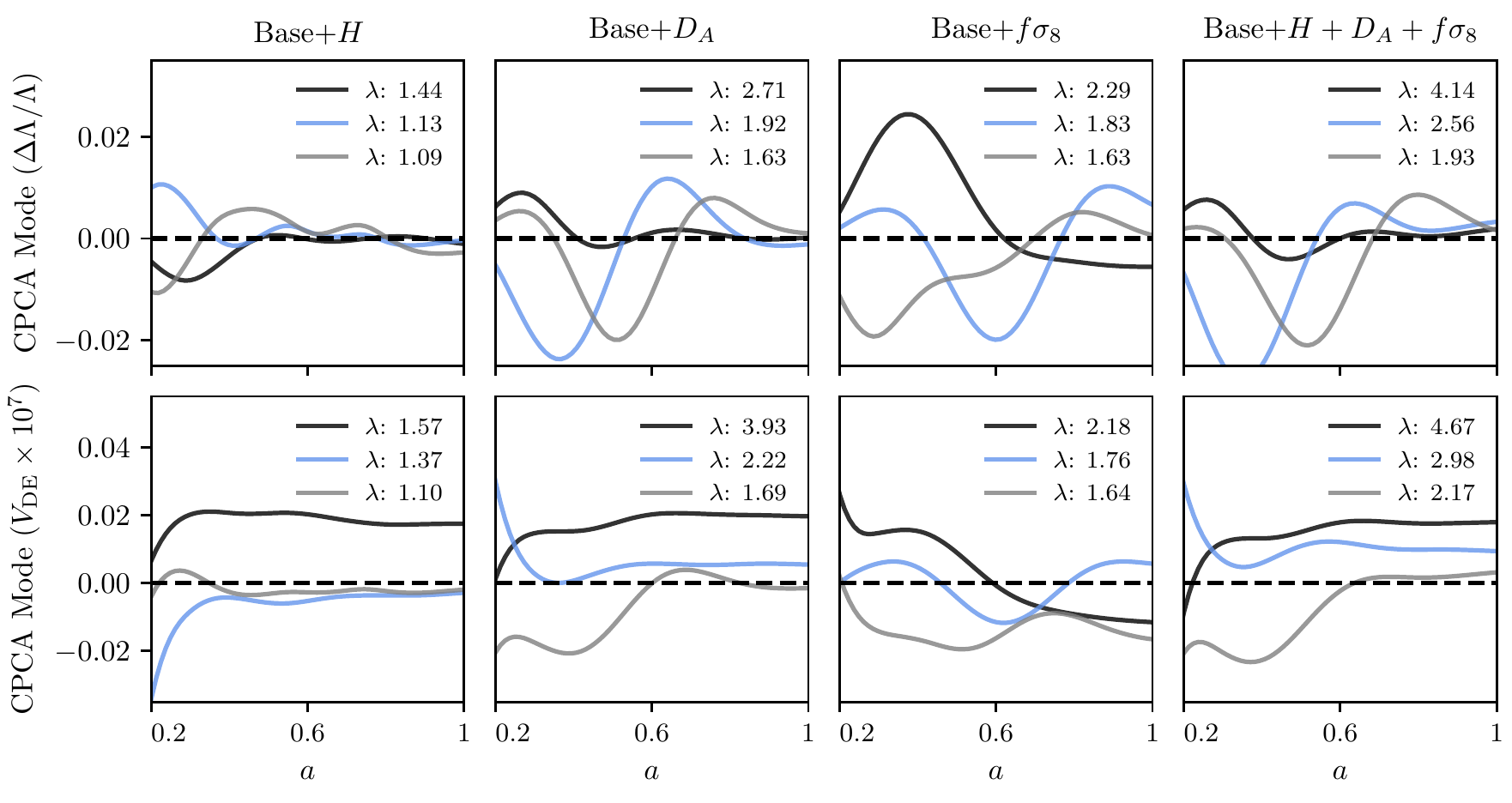}
\caption{\textit{Top}: the three best constrained CPCA modes of the reconstructed EFT function for combinations of the baseline dataset with DESI quintessence mocks. \textit{Bottom}: the three best constrained CPCA modes of the quintessence potential for combinations of the DESI quintessence datasets.  The corresponding CPCA eigenvalues are shown in the legend.}
\label{fig:cpca_mode_EFT}
\end{figure*}

In Fig.~\ref{fig:cpca_mode_EFT} we plot the three most constrained CPCA modes for the DESI quintessence analyses of the reconstructed EFT function (top) and reconstructed potential (bottom). We first focus on the results of the reconstructed potential. Recall that for the EFT reconstruction of the DESI $\Lambda$CDM mock datasets, the best constrained mode of the reconstructed potential for datasets that included mock DESI BAO measurements was constant and highly correlated with $H_0$ and $\Omega_m$. This mode resulted from improved constraints on $H_0$ and $\Omega_m$ after including measurements of $H(z)r_s$ and $D_A(z)/r_s.$ For the quintessence mocks, the two best constrained modes of the analyses including BAO data are constant modes. We again find that the first mode is highly correlated (70-80\%) with $H_0$ and $\Omega_m$, hence it corresponds to the same rescaling of the amplitude induced by improved constraints on $H_0$ and $\Omega_m$. On the other hand, the second constant mode is largely uncorrelated ($<20\%$) with all $\Lambda$CDM parameters. In effect, the second mode corresponds to a rescaling of the amplitude that cannot be explained by simply shifting the $\Lambda$CDM parameters. This mode arises because the baseline dataset is in tension with the quintessence mock that we generated, hence we must realize a non-zero EFT function to appropriately account for this tension. This also explains why the angular diameter distance measurements constrain $\Delta\Lambda/\Lambda$ for the quintessence analysis, but not for the $\Lambda$CDM analysis. For the $\Lambda$CDM mock datasets, the information added by the angular diameter distance could be fully described by shifting the $\Lambda$CDM parameters, but this is not the case for the quintessence mock. The angular diameter distance measurements also have a more significant contribution to the CPCA modes of $\Delta \Lambda/\Lambda$ for the full quintessence analysis than they did for the $\Lambda$CDM datasets. Moreover, the best constrained mode of the angular diameter distance measurements corresponds to the best constrained mode of the fiducial analysis. The ability to distinguish between the sources of these constraints demonstrates the usefulness of CPCA in revealing details about what a dataset is actually constraining.

Whereas including DESI data leads to significant improvements in the constraints on the amplitude of the quintessence potential, the overall shape remains relatively unconstrained. As a test, we search for the existence of a local minima within one correlation length of the mocks actual local minimum ($z\in [0.2,1.1]$) and find that the fraction of samples possessing a minimum in this interval does not increase after adding the DESI mocks. This is not surprising given that our EFT reconstruction technique explicitly smooths out features on sub-correlation length scales in order to avoid overfitting the data. Using alternative methods to constrain quintessence, such as those that require an explicit functional form for the quintessence potential, will likely be able to better detect localized features in the potential. 

Analysis of further quintessence properties such as the dark energy equation of state and Swampland related observables provides the same conclusions as those reached in Sec.~\ref{Sec:quint_improvement}, hence we do not analyze those results here.

\section{Conclusions} \label{Sec:Conclusions}

In this work we present forecasted constraints on $\Lambda$CDM and single field quintessence models for a five year DESI survey. Using current CMB, LSS, and SN datasets as a baseline, we find that including mock DESI measurements of the Hubble parameter, angular diameter distance, and linear growth rate will significantly improve constraints on $\Lambda$CDM and quintessence cosmologies. We list our main findings below:

\begin{itemize}
    \item Combined analysis of $H(z)r_s$, $D_A(z)/r_s$, and $f\sigma_8(z)$ improves constraints on $H_0,$ $\Omega_m$, and $\sigma_8$ by a factor of two with the best constrained linear combination of parameters corresponding to a combination of $H_0$ and $\sigma_8.$  Constraints on $\sigma_8$ come entirely from $f\sigma_8(z)$ and constraints on $H_0$ and $\Omega_m$ come primarily from $D_A(z)/r_s$. Including only the Hubble parameter in our $\Lambda$CDM analysis does not improve our constraints on $\Lambda$CDM parameters.
    \item  Our quintessence reconstruction favors lower values of $H_0$ and $\sigma_8$ and higher values of $\Omega_m$ than our $\Lambda$CDM analyses. These differences indicate the importance of viewing constraints on cosmological parameters within the context of the assumed cosmology. 
    \item BAO ($H(z)r_s$ and $D_A(z)/r_s$) and RSD measurements ($f\sigma_8(z)$) are sensitive to different aspects of the quintessence potential. In particular, BAO measurements predominately constrain the amplitude of the potential by constraining $H_0$ and $\Omega_m$. On the other hand, the growth of structure constraints on the potential are largely independent of the constraints on the $\Lambda$CDM parameters, and instead arise from improved constraints on the EFT function $\Delta \Lambda/\Lambda.$ In the presence of a quintessence signal which cannot be accounted for by shifting of the cosmological parameters, the BAO measurements can constrain the amplitude of the potential potential via $H_0$ and $\Omega_m$, as well as by constraining the reconstructed EFT function.
    \item Constraints on the dark energy equation of state and its time evolution improve significantly after including mock DESI measurements of $f\sigma_8(z).$ Should DESI observe a Universe consistent with $\Lambda$CDM, then the improved constraints on $w_{\rm DE}$ and its phase space will significantly restrict the late time dynamical behavior of viable quintessence models.
    \item Including DESI mocks does not improve constraints on $|\nabla_\phi V|/V$ and $-\nabla^2_\phi V/V$, hence it is unlikely that we will be able to test the Swampland Conjectures with more precision than the baseline analysis \cite{Park:2021qnt} using the EFT reconstruction on DESI data. Nevertheless, including DESI data, particularly $f\sigma_8(z)$, leads to significant improvements on constraints of the scalar field excursion.

    \item Our EFT reconstruction of mock DESI datasets is unable to identify localized features in the potential. Identifying such features for quintessence cosmologies will likely require alternative modeling techniques or more constraining data. 
    \item Analyses of each DESI observable individually suggest that most of our constraints come from $D_A(z)/r_s$ and $f\sigma_8(z)$. Since the precision of growth rate measurements depends heavily on our non-linear modelling capabilities, these results emphasize the importance of developing robust techniques to model non-linear RSD for biased tracers in order to extract optimal cosmological constraints from Stage-IV galaxy surveys \cite{Perko_2016, Chen_2021}.
\end{itemize}

The goal of this work was not only to forecast constraints on $\Lambda$CDM and quintessence cosmologies for DESI, but also to demonstrate the usefulness of the CPCA metrics in interpreting cosmological constraints. The analyses considered here, particularly the EFT reconstruction of quintessence cosmologies, relied heavily on the CPCA techniques to answer the question of what do our datasets actually measure? In decomposing the posterior covariances into their CPCA modes and eigenvalues we are able to determine the which parameter constraints are associated with a given dataset and, perhaps more importantly, which parameter constraints are a consequence of constraints on alternative, often degenerate parameters.

We emphasize that our results are specific to the EFT reconstruction technique and summary statistics used in this work. More advanced modelling techniques such as incorporating the full shape of the power spectrum, non-Gaussian statistics, and including more small scale information could provide better constraints on $\Lambda$CDM and modified gravity/dark energy models as discussed in \cite{Alam_2021}. Nevertheless, we still find significant improvements in both $\Lambda$CDM and quintessence constraints when considering only BAO distance scale and linear growth rate measurements.

The most immediate follow up to this project would be to perform the analysis using actual DESI observations once those become available. Additionally one could use the methods presented here to forecast constraints on more general theories of modified gravity and dark energy, beyond single field, minimally coupled quintessence. Finally, a comparison between the forecasts of this work with those using more advanced modelling techniques could help reveal details in how to optimally extract information from large volume galaxy surveys.

\acknowledgments

SG acknowledges support from the Fulbright U.S. Student Program.
MR is supported in part by NASA ATP Grant No. NNH17ZDA001N, and by funds provided by the Center for Particle Cosmology. 
BJ is supported in part by the US Department of Energy Grant No. DE-SC0007901.  
LS is grateful for the support from the Shota Rustaveli NSF of Georgia (grant FR/18-1462) and DOE grants DE-SC0021165 and DE-SC0011840, and the NASA ROSES grant 12-EUCLID12-0004. Computing resources are provided by the University of Chicago Research Computing Center through the Kavli Institute for Cosmological Physics at the University of Chicago. 

\bibliographystyle{apsrev4-1}
\bibliography{biblio}

\end{document}